\documentclass[prd,twocolumn,showpacs,preprintnumbers,amsmath,amssymb,superscriptaddress,nofootinbib,english]{revtex4-1}
\usepackage{graphicx,natbib,times}
\usepackage{url}
\usepackage{color}
\usepackage{rotating}
\usepackage{hyperref}
\usepackage{mathrsfs}
\usepackage{enumitem}
\usepackage{hyperref}
\usepackage{cleveref}
\hypersetup{
    colorlinks=true,
    linkcolor=red,
    citecolor=blue,
}

\newcommand{\mnras}{{Mon.~Not.~R.~Astron.~Soc.}}
\newcommand{\aap}{{A\&A}}

\newcommand{\jcap}{{J. Cosmol. Astropart. Phys.}}

\def\be{\begin{equation}}
\def\ee{\end{equation}}
\def\ba{\begin{eqnarray}}
\def\ea{\end{eqnarray}}
\def\gsim{\;\rlap{\lower 2.5pt
\hbox{$\sim$}}\raise 1.5pt\hbox{$>$}\;}
\def\lsim{\;\rlap{\lower 2.5pt
\hbox{$\sim$}}\raise 1.5pt\hbox{$<$}\;}

\begin{document}

\title{Neutrino effects on the morphology of cosmic large-scale structure}

\author{Yu Liu}
\email{liu-yu@sjtu.edu.cn}
\affiliation{Department of Astronomy, School of Physics and Astronomy, Shanghai Jiao Tong University, Shanghai, 200240, P. R. China}

\author{Yu Yu}
\email{Corresponding author.\\yuyu22@sjtu.edu.cn}

\affiliation{Department of Astronomy, School of Physics and Astronomy, Shanghai Jiao Tong University, Shanghai, 200240, P. R. China}

\author{Hao-Ran Yu}
\email{haoran@xmu.edu.cn}

\affiliation{Department of Astronomy, Xiamen University, Xiamen, Fujian 361005, P. R. China}
\affiliation{Tsung-Dao Lee institute, Shanghai, 200240, P. R. China}
\affiliation{Department of Astronomy, School of Physics and Astronomy, Shanghai Jiao Tong University, Shanghai, 200240, P. R. China}

\author{Pengjie Zhang}
\email{zhangpj@sjtu.edu.cn}

\affiliation{Department of Astronomy, School of Physics and Astronomy, Shanghai Jiao Tong University, Shanghai, 200240, P. R. China}
\affiliation{Tsung-Dao Lee institute, Shanghai, 200240, P. R. China}
\affiliation{Shanghai Key Laboratory for Particle Physics and Cosmology, Shanghai 200240, P. R. China}

\begin{abstract}
In this work, we propose a powerful probe of neutrino effects on the large-scale structure (LSS) of the Universe, i.e., Minkowski functionals (MFs). The morphology of LSS can be fully described by four MFs. This tool, with strong statistical power, is robust to various systematics and can comprehensively probe all orders of N-point statistics. By using a pair of high-resolution N-body simulations, for the first time, we comprehensively studied the subtle neutrino effects on the morphology of LSS. For an ideal LSS survey of volume $\sim1.73$ Gpc$^3$/$h^3$, neutrino signals are mainly detected from void regions with a significant level up to $\thicksim 10\sigma$ and $\thicksim 300\sigma$ for CDM and total matter density fields, respectively. This demonstrates its enormous potential for much improving the neutrino mass constraint in the data analysis of up-coming ambitious LSS surveys.
\end{abstract}

\pacs{ }

\maketitle
 
\section{Introduction} 
Neutrino mass problem is one of major challenges in fundamental physics. The $Z$ boson lifetime measurements found that the number of active neutrinos is 3 ($N^{active}_{\nu}$ = $2.9840\pm0.0082$) \cite{2006PhR...427..257A}, and the neutrino oscillation experiments also revealed that at least two of the three neutrino eigenstates are massive \cite{1992PhRvD..46.3720B, 1998PhRvL..81.1562F, 2004PhRvL..92r1301A}. However, the oscillation experiments only give the mass-squared splittings between the neutrino eigenstates, which implies lower bound on the sum of neutrino masses, $\Sigma m_{\nu}$, to be 0.05 and 0.1 eV for the normal and inverted-mass hierarchies (e.g., \cite{2008PhRvL.101m1802A}), respectively. The beta decay and neutrinoless double-beta decay experiments are the promising laboratory-based experiments for obtaining the absolute neutrino mass scale. Nevertheless, due to current technical limitations in particle physics experiments (e.g., \cite{2010NIMPA.623..442W, 2017JPhG...44e4004A}), further accurate measurement of absolute neutrino mass will be challenging.

In cosmology, the analysis of cosmological observables [e.g., anisotropies of cosmic microwave background (CMB) and distribution of LSS] can provide crucial complementary information on neutrino masses beyond particle physics experiments. At present, the strongest constraint on the upper bound of neutrino mass sum, $\Sigma m_{\nu} < 0.12$ eV (2$\sigma$), comes from cosmology by combination analysis of CMB and BAO data assuming $\Lambda$CDM cosmology \cite{2018arXiv180706209P}. The next-generation LSS surveys (e.g., SKA \cite{SKA},
DESI \cite{2016arXiv161100036D}, LSST \cite{2009arXiv0912.0201L}, WFIRST \cite{wfirst}, Euclid \cite{euclid}) and CMB surveys (e.g., the Simons Observatory \cite{2019JCAP...02..056A} and CMB-S4 \cite{2016arXiv161002743A}) will map the cosmic large-scale structure with high precision, which provides great opportunity to improve the measurements of neutrino mass sum upper bound and other cosmological parameters.

Cosmic neutrinos with large thermal velocities can suppress the density perturbations below their free streaming scale, $\lambda_{fs}(m_{\nu},z) = a(2\pi/k_{fs}) \simeq 7.7(1+z)/[\Omega_{\Lambda}+\Omega_{m}(1+z)^3]^{1/2}$($1$eV)/$m_{\nu}$ Mpc/$h$ \cite{1998PhRvL..80.5255H, article, 2011ARNPS..61...69W, 2013neco.book.....L}. The damping amplitude of density perturbation on nonlinear scales depends on the total neutrino masses, which has been commonly used to constrain and forecast the $\Sigma m_{\nu}$ (e.g., \cite{1998PhRvL..80.5255H, 2008PhRvL.100s1301S, 2016MNRAS.462.4208P, 2017JCAP...02..052A, 2019arXiv190706666C}). In linear theory, the damping amplitudes, $|\Delta P/P|$, on small scales, $k\lambda_{fs}\gg1$, in total matter power spectrum and in CDM power spectrum are $\sim 8f_{\nu}$ and $\sim 6f_{\nu}$, respectively \cite{2019arXiv190706598B}. Here, the neutrino mass fraction is defined by $f_{\nu} \equiv \Omega_{\nu}/\Omega_{m}$, and density parameter of non-relativistic neutrinos is given by $\Omega_{\nu}=\Sigma m_{\nu}/(93.14h^2$eV) \cite{article}. On large scales, $k\lambda_{fs}\ll1$, neutrinos cluster just as CDM and baryonic matter.

However, the damping level on power spectrum (two-point statistics) is small for realistic neutrino masses, $f_{\nu} \lesssim 1\%$, which makes the damping effect easily contaminated by uncertainties from different sources, e.g., non-linear bias, redshift space distortions (RSDs), baryonic effects \cite{2019JCAP...01..010P} and degeneracies with $\sigma_8$ \cite{2018ApJ...861...53V}. Worse still, two-point statistics can only capture Gaussian information, missing substantial higher-order information for density field being highly non-Gaussian at late Universe, while neutrino signals are basically detected around nonlinear scales. These deficiencies downgrade their power for neutrino mass constraints. Other possible unknown systematics beyond standard $\Lambda$CDM cosmology may also mimic neutrino effect on matter power spectrum and consequently affect neutrino mass constraints (e.g., nonzero curvature, dynamical dark energy, modified gravity \cite{2017PhRvL.118r1301F, 2019A&A...629A..46H, 2019JCAP...06..040W}, interactions in the dark sector, etc.). For these reasons, there is strong motivation to investigate new neutrino effects (e.g., \cite{2019arXiv190500361Z, 2019PhRvD..99l3532Y}) and novel alternative methods beyond two-point statistics (e.g., \cite{2018JCAP...03..003R, 2019JCAP...05..043C, 2019PhRvD..99f3527L, 2019JCAP...06..019M}). At meanwhile, accurate modeling of neutrino effects is also becoming increasingly essential and critical to the neutrino study in cosmology.
 
In this work, we propose a powerful non-Gaussian probe of neutrino effects on LSS, i.e., Minkowski functionals (MFs), toward improving constraining power on $\Sigma m_{\nu}$ in data analysis of up-coming LSS surveys. This method can comprehensively capture all orders of N-point statistics \cite{2017PhRvL.118r1301F} of LSS and be robust to various systematic effects \cite{Park_2010, 2001MNRAS.327.1041B, 2012ApJ...747...48W, 2014MNRAS.437.2488B, 10.1093/pasj/55.5.911, 2017PhRvL.118r1301F}, e.g., nonlinear evolution, nonlinear bias, RSDs, etc. In particular, its potential in constraining $\Sigma m_{\nu}$ was only addressed for the $2D$ weak lensing (WL) convergence field in Ref. \cite{2019JCAP...06..019M}, where the neutrino effects on WL correspond to that on the projected LSS (along line of sight) in between source and observer. In this work, we mainly focus our study on the analysis of neutrino effects on LSS, by using $3D$ MFs. In comparison with previous case-by-case studies (e.g., neutrino impacts on voids \cite{2015JCAP...11..018M, 2019MNRAS.488.4413K} and halos/clusters \cite{2012PhRvD..85f3521I, 2013JCAP...12..012C}, which can only capture local information of neutrino effects on LSS), analysis by using MFs is helpful to comprehensively understand subtle neutrino effects on different density regions of LSS. Moreover, we find neutrino signals in MFs are mainly detected from underdense regions, which makes the neutrino detections potentially avoid various systematics from high density regions. Due to including higher-order information, non-Gaussian tools (e.g., $2D$ MFs \cite{2012PhRvD..85j3513K, 2017MNRAS.466.2402S}, peak statistics \cite{2017MNRAS.466.2402S, 2018A&A...619A..38P}, three-point statistics \cite{2017MNRAS.466.2402S, 2010APh....32..340V, 2019arXiv190911107H}, etc. \cite{2018A&A...619A..38P, 2019A&C....27...34X}) combined with other probes also help breaking parameter degeneracies in various cosmological studies.

\section{Minkowski functionals}
Minkowski Functionals are a set of morphological descriptors. They are all additive, motion invariant, which makes them insensitive to observational effects, e.g., the survey shape \cite{10.1093/pasj/55.5.911}. This tool, originally derived from theory of convex bodies and integral geometry, was first introduced to cosmology by Ref. \cite{1994A&A...288..697M}, and then was commonly used to detect deviations from Gaussianity (e.g., \cite{2013MNRAS.429.2104D, 2016A&A...594A..17P}). According to Hadwiger's theorem \cite{Hadwiger1957Vorlesungen}, the morphological properties of any pattern in $d$-dimensional space can be fully characterized by $d+1$ MFs, which allows MFs to comprehensively probe all orders of N-point statistics at once. Therefore, MFs can be served as a powerful non-Gaussian statistical tool in cosmology to provide extra information beyond popular two-point statistics, leading to improving power on cosmological parameter constraint (e.g. $\Omega_m$, $\sigma_8$, $w$ and $\Sigma m_{\nu}$ in $2D$ weak lensing convergence field analysis \cite{2012PhRvD..85j3513K, 2019JCAP...06..019M}).

For $3D$ LSS analysis in cosmology, the most commonly used patterns (other patterns also could be found in literatures, e.g., \cite{1994A&A...288..697M}) are the excursion sets ($F_{\nu}$) of matter density field (or halo/galaxy field), where the density threshold ($\nu$) is adopted to be diagnostic parameter for displaying the morphological features. Here, the excursion set $F_{\nu}$ is the set of all points $\mathbf{x}$ with density $\nu(\mathbf{x})\geq \nu$. The Minkowski Functionals measure the volume ($V_0$) and the surface's area ($V_1$), integrated mean curvature ($V_2$), and Euler characteristic ($V_3$) of the excursion set, normalized by the whole field volume $|\mathscr{D}|$,
\begin{equation}\label{1}
\begin{aligned}
    &V_{0}(\nu)=\frac{1}{|\mathscr{D}|}\int_{F_{\nu}}d^3x,\\
    &V_{1}(\nu)=\frac{1}{6|\mathscr{D}|}\int_{\partial F_{\nu}}dS(\mathbf{x}),\\
    &V_{2}(\nu)=\frac{1}{6\pi|\mathscr{D}|}\int_{\partial F_{\nu}}(\frac{1}{R_1(\mathbf{x})}+\frac{1}{R_2(\mathbf{x})})dS(\mathbf{x}),\\
    &V_{3}(\nu)=\frac{1}{4\pi|\mathscr{D}|}\int_{\partial F_{\nu}}\frac{1}{R_1(\mathbf{x})R_2(\mathbf{x})}dS(\mathbf{x}),\\
\end{aligned}
\end{equation}
where $R_1(\mathbf{x})$ and $R_2(\mathbf{x})$ are the principal radii of curvature of the excursion set's surface orientated toward lower density region. The first two MFs describe the size of the excursion set, and the last two MFs characterize the shape (geometrical property) and connectivity (topological property) of the set surface (isodensity contours at level $\nu$), respectively. The last MF is simply related to the genus ($G=1-V_3$), that is the first topological descriptor commonly used in cosmology (e.g., \cite{1986ApJ...306..341G, 2012ApJ...747...48W} ). The topological Euler characteristic $\chi$, obtained through a surface integration of the Gaussian curvature according to the Gauss-Bonnet theorem, is proportional to $V_3$ by a factor 2, $\chi=2V_3$. And, $V_3$ is related to the number of isolated regions (balls) with density above a given threshold, empty regions inside balls (bubbles) and holes in ball surfaces (tunnels) per unit volume, $V_3=\frac{1}{|\mathscr{D}|}(N_{\rm{ball}}+N_{\rm{bubble}}-N_{\rm{tunnel}})$. This makes it more convenient to use than $G$ due to its additivity. Moreover, it is also insensitive to systematic effects \cite{Park_2010, 2001MNRAS.327.1041B, 2012ApJ...747...48W, 2014MNRAS.437.2488B}, since the intrinsic topology can be well conserved during deformation.

There are two standard numerical methods (i.e., the Koenderink invariant and the Crofton's formula) developed by \cite{Schmalzing_1997} for measuring density field's MFs. The MFs of Gaussian random field have analytic expressions, which remarkably agree with these numerical results \cite{MELOTT19901, Schmalzing_1997}. In this work, we choose the Crofton's formula method to quote our results, for the two methods giving consistent results.

\section{N-body simulations}
Beyond the attempts to understand neutrino effects on LSS analytically (e.g., \cite{2012PhRvD..85f3521I, 2015JCAP...03..046F}), the neutrino cosmological N-body simulations are essential to study neutrino nonlinear dynamics. Various approaches have been proposed to implant massive neutrinos into the standard N-body simulations, e.g., the particle-based, the grid-based \cite{2009JCAP...05..002B}, the linear response \cite{2013MNRAS.428.3375A}, the hybrid approach between the particle-based and the grid-based \cite{2010JCAP...01..021B} (or the linear response \cite{2018MNRAS.481.1486B}) and even fluid techniques \cite{2016JCAP...11..015B, 2017PhRvD..95f3535I}. In general, the grid-based and the linear response approaches cannot accurately resolve the non-linear neutrino structure formation on small scales, which can be alleviated by the hybrid approaches. While, particle-based approach can naturally capture the full non-linear neutrino clustering. But at meanwhile, this method is hindered by Poisson noise on small scales (induced by the large thermal motion of neutrinos), which has to be reduced by increasing the number of neutrino particles in the simulation.
Our neutrino N-body simulation (\emph{TianNu}) adopt the particle-based approach. For reducing Poisson noise, \emph{TianNu} incorporates neutrinos with pushing to the extreme scales, which makes it currently one of world's largest cosmological N-body simulations \cite{2017RAA....17...85E}.

Specifically, we adopt a pair of high-resolution N-body simulations (i.e., \emph{TianZero} with $\Sigma m_{\nu}=0$ eV and \emph{TianNu} with $\Sigma m_{\nu}=0.05$ eV \cite{2017RAA....17...85E}) realized using publicly-available code, CUBEP3M \cite{2013MNRAS.436..540H}, for resolving the subtle neutrino effects between neutrinos and CDM, especially on non-linear scale \cite{2017PhRvD..95h3518I, 2017NatAs...1E.143Y}. CUBEP3M here is optimized using hybrid-parallelized Particle-Mesh (PM) algorithm for long-range gravitational force calculation, plus an adjustable Particle-Particle (PP) algorithm ($r_{soft} = L/(20n_{p}^{1/3})$) for increasing resolution below mesh scale. Both simulations were initialized at $z = 100$ with the same initial condition parameterized with [$\Omega_c$, $\Omega_b$, $h$, $n_s$, $\sigma_8$] = [$0.27$, $0.05$, $0.67$, $0.96$, $0.83$], evolving $n_p = 6912^3$ CDM particles with mass resolution of $7\times10^8 M_\odot$ in periodic cubic box of width $L = 1200$ Mpc/$h$ (volume $\sim1.73$ Gpc$^3$/$h^3$). In \emph{TianNu}, $13824^3$ neutrino particles with mass resolution of $3\times10^5 M_\odot$ are incorporated into the mixture with $\Omega_m$ fixed for cleanly extracting neutrino effects. Here, the minimal normal hierarchy mass model is chosen to simulate neutrinos with one massive species ($m_\nu = 0.05$ eV) treated as particles and other two light species ($m_\nu = 0$ eV) included in background cosmology by using the CLASS \cite{2011JCAP...07..034B} transfer function.

\section{Data}
Analysis and results in this work are based on density fields at $z = 0.01$, which is instrumental in forecasting neutrino signatures from a shallower, lower-redshift galaxy survey with high number density (e.g., Bright Galaxy Survey (BGS) sample within $0.05 < z < 0.4$ in DESI \cite{2016arXiv161100036D}). Here, the advantage of using density fields to perform analysis is that it can help us better understand subtle neutrino effects on LSS. Both CDM field ($\Phi_{dm}$ in \emph{TianZero} and \emph{TianNu}) and total matter field ($\Phi_{total}$ in \emph{TianNu}) are computed by Cloud-In-Cell (CIC) interpolation technique onto $N_g = 2048^3$ regular grids. For interpolation of $\Phi_{total}$ in \emph{TianNu}, each particle is weighted by a factor of $\Omega_i/(\Omega_m N_i)$, where $\Omega_i$ and $N_i$ are the energy fraction and number of particles of species $i$, repectively. We subsequently smooth these fields separately by two Gaussian window functions with different smoothing scales, $R_G$ (i.e., $0.2L_g = 0.12$ Mpc/$h$ and $0.4L_g = 0.24$ Mpc/$h$, where $L_g \equiv L/N_g^{1/3}$ is the grid size), to obtain the smoothed fields. These Gaussian smoothed fields serve for investigating the impacts of smoothing on our results. The MFs are then measured for all these fields as a function of $\rho/\overline{\rho} \equiv 1 + \delta$, which is the density threshold used to define the excursion set. We compare the MFs measured from different cosmology models (i.e., $\Lambda$CDM and $\nu\Lambda$CDM) to highlight neutrino signatures and analyze the neutrino effects on LSS.

\begin{figure*}[ht]
\hspace*{-0.7cm}
{\includegraphics[angle=0,scale=1.251,trim=0cm 0cm 0cm 0cm,clip=true]{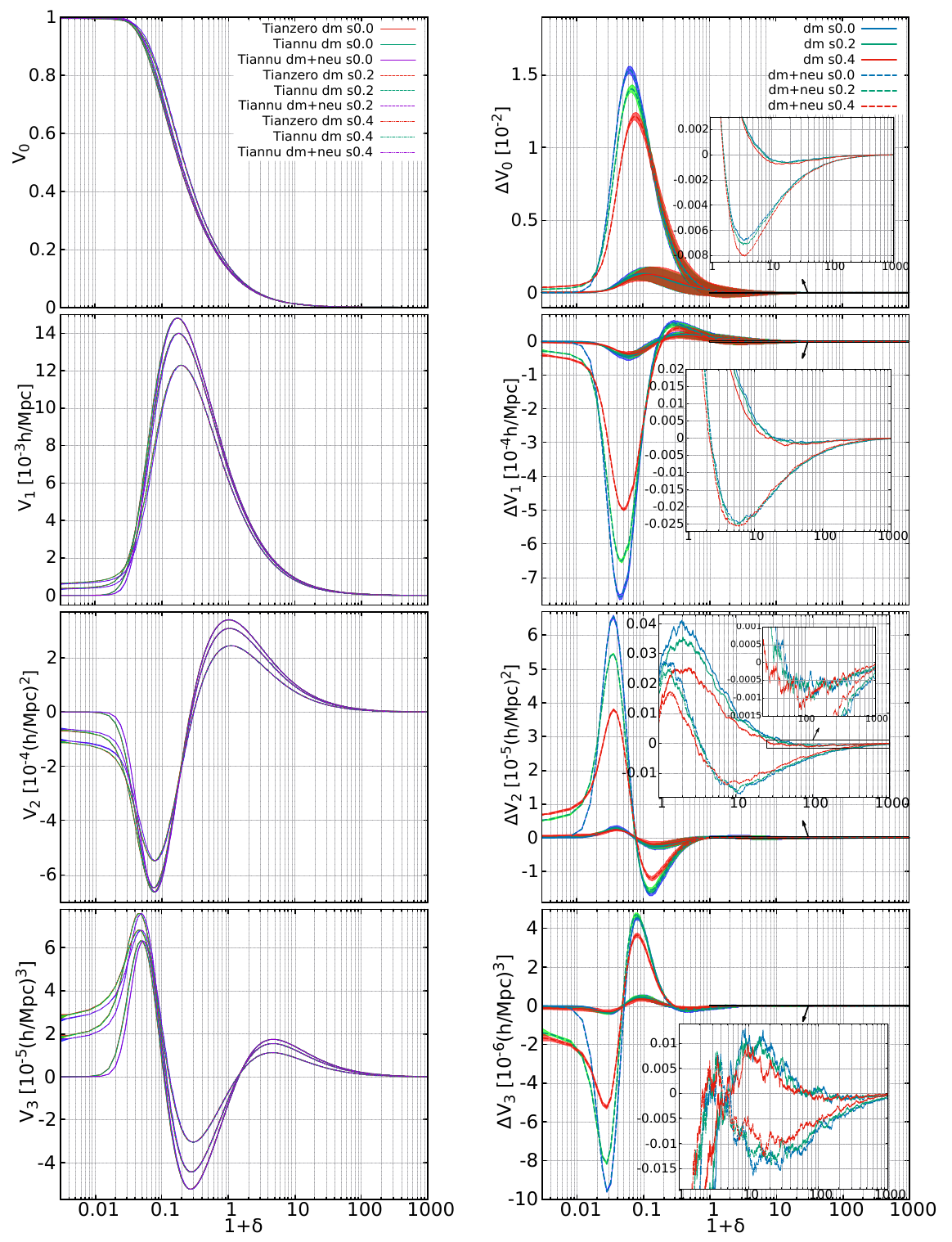}}
\caption{Left panels: The MFs of LSS as functions of $1 + \delta$, which are computed from the density fields with different smoothing scales, i.e., $0L_g$, 0.2$L_g$, 0.4$L_g$, at $z = 0.01$. For \emph{TianNu}, the MFs of CDM fields and total matter fields are measured. Here, the error bars are too tiny to be visible. The $1 + \delta$, i.e., $\rho/\overline{\rho}$, is the density threshold used for the calculations of MFs. Right panels: The differences in the MFs between $\Lambda$CDM and $\nu\Lambda$CDM cosmology ($\Sigma m_{\nu} = 0.05$ eV). The error bars in subpanels are omitted for clarity.}
\label{fig:mf1}
\end{figure*}

\section{Neutrino effects on the morphology of LSS}
Our results are presented in Figure~\ref{fig:mf1}. Left panels show the MFs themselves, while the differences in MFs between $\nu\Lambda$CDM and $\Lambda$CDM cosmology, the $\Delta V_i$s, are displayed in right panels. The results are well visualized by logarithmic x-axis in the range of [0.003, 1000], considering that the probability distribution function of density field roughly obeys lognormal form at low redshift \cite{1991MNRAS.248....1C}. The error bars are estimated by the standard errors \cite{numerical.book} of MFs of $\Phi_{dm/total}$ (i.e., $\Phi_{dm}$ or $\Phi_{total}$), $s_e = \sigma / \sqrt{n}$, where the $\sigma$ is the standard deviation of the MFs measured from $n$ ($8^3 = 512$) sub-fields ($L_{sub} = 150$ Mpc$/h$) obtained by equal-subdivided $\Phi_{dm/total}$. The $\Delta V_i$s are measured by two cases, i.e., $\Delta V_i^{dm/total} \equiv V_i(\Phi_{dm/total})_{\rm{TianNu}} - V_i(\Phi_{dm})_{\rm{TianZero}}$, considering that $\Phi_{dm}$ and $\Phi_{total}$ can in principle be inferred from galaxy clustering and weak lensing \cite{2007Natur.445..286M} (or integrated Sachs-Wolfe effect) from various cosmological surveys, respectively. In the following, the neutrino effects on LSS are resolved by understanding the $\Delta V_i$s. Nevertheless, we will also mention $V_i$s when they are necessary for helping our understanding of $\Delta V_i$s.

In linear theory, cosmic neutrino background can slow down the growth of CDM perturbations, e.g., on scales $k \gg k_{nr} \simeq 0.018(\Omega_m)^{1/2}(m_{\nu}/1$eV$)^{1/2}$ $h$Mpc$^{-1}$, the $\delta_{dm} \propto a$ is replaced by $\delta_{dm} \propto a^{p_+} \simeq a^{1-3/5f_{\nu}}$ during matter domination and $\delta_{dm} \propto ag(a)$ is replaced by $\delta_{dm} \propto [ag(a)]^{p_+} \simeq [ag(a)]^{1-3/5f_{\nu}}$ during $\Lambda$ domination, where $g(a)$ is a damping factor normalized to $g = 1$ for $a \ll a_{\Lambda}$  \cite{article} (corresponding to the global slowdown of structure growth caused by $\Lambda$). Overall, in Figure~\ref{fig:mf1} the $\Delta V_i$s measured by the two cases have the same trend, which can be well interpreted by the aforementioned neutrino effects. For \emph{TianNu}, $\Phi_{total}$ is partially contributed by neutrinos, i.e., $\delta_{total} = f_{dm}\delta_{dm} + f_{\nu}\delta_{\nu}$, where $f_{dm} \equiv \Omega_c/\Omega_m$ and $f_{\nu} \approx 0.37\%$. While, the clustering of neutrinos is much weaker than that of CDM, due to neutrino free-streaming ($\lambda_{fs}(0.05$ eV$,0.01) \approx 150$ Mpc/$h$). Therefore, the matter perturbations in $\Phi_{total}$ are slightly lower than that in $\Phi_{dm}$, which makes the amplitudes of $\Delta V_i^{total}$s are relatively larger than that of $\Delta V_i^{dm}$s (cf. Figure~\ref{fig:mf1}).

When $\rho/\overline{\rho}$ is low enough, the complement of excursion set will be the isolated void regions with closed surfaces whose positive directions point inward, which leads to a negative mean curvature ($\overline{K}$) of the excursion set's surface, i.e., $V_2 < 0$. Specifically, in the range of $\rho/\overline{\rho} \lesssim 0.2$, we find the $\Delta V_0 > 0$ and $\Delta V_1 < 0$, which means that voids' sizes become smaller and their inner matter becomes denser with presence of massive neutrinos \cite{2015JCAP...11..018M}. Therefore, the mean curvature ($\overline{K}$) of the excursion set's surface at meanwhile becomes smaller, i.e., $\Delta\overline{K} < 0$. These results can be well understood, since neutrinos contribute to the interior mass of underdense regions and slow down CDM evacuation from voids \cite{2015JCAP...11..018M}. The trend of $\Delta V_2$ in this range is a little bit complicated, since $\Delta V_2$ is the combination result between $\Delta V_1$ and $\Delta \overline{K}$. We note that $V_2$ can be roughly expressed by $ V_2\sim\overline{K} \cdot V_1$ ($\Delta V_2 \sim \Delta \overline{K} \cdot V_1 + \overline{K} \cdot \Delta V_1$), where $V_1$ is always positive \cite{2017PhRvL.118r1301F}. Therefore, when dominated by $\Delta V_1$, $\Delta V_2$ follows a completely opposite trend with $\Delta V_1$ in the range of $\rho/\overline{\rho} \lesssim 0.08$ considering $\overline{K} < 0$, i.e., $\Delta V_2 > 0$; when dominated by $\Delta \overline{K}$, $\Delta V_2$ shares the same trend with $\Delta\overline{K}$ in the range of $0.08 \lesssim \rho/\overline{\rho} \lesssim 0.2$, i.e., $\Delta V_2 < 0$.

For a higher $\rho/\overline{\rho}$, the excursion set will turn into the non-virialized web-like skeletons surrounding voids, which makes a positive $\overline{K}$, i.e., $V_2 > 0$. The transition from negative to positive of $V_2$ happens at $\rho/\overline{\rho} \in [0.2, 0.3]$ in our study. Here, we note that the accurate $V_i$s depend on the smoothing ($R_g$) and resolution ($N_g$) of the density field, which can result in different transition point of $V_2$. In the range of $0.2 \lesssim \rho/\overline{\rho} \lesssim 1$, we find $\Delta V_0 > 0$, $\Delta V_1 > 0$ and $\Delta V_2 < 0$, since neutrino background delays the structure growth making web-like skeletons bigger and looser. Here, the $\overline{K}$ still becomes smaller, i.e., $\Delta\overline{K} < 0$, and $\Delta V_2$ is dominated by $\Delta\overline{K}$. When $\rho/\overline{\rho}$ is high enough, for the same reason, the over-dense regions ($\rho/\overline{\rho} \gtrsim 1$) shrink in size, making the excursion set smaller and resulting in $\Delta\overline{K} > 0$. Therefore, we see that the $\Delta V_0$ and $\Delta V_1$ transit from positive to negative in the range of $\rho/\overline{\rho} \gtrsim 1$. Meanwhile, we find $\Delta V_2 > 0$ in the vicinity of $\rho/\overline{\rho} \approx 1$, where $\Delta\overline{K}$ plays a key role in $\Delta V_2$. When $\rho/\overline{\rho}$ reaches a high enough level ($\rho/\overline{\rho} \gg 1$), $\Delta V_2$ will be dominated by $\Delta V_1$, making them share the same trend, i.e., $\Delta V_2 < 0$.

For understanding $\Delta V_3$, we need deep insights in hierarchical void formation, since topology of the excursion set heavily relies on the subtle structures of LSS. For void hierarchy \cite{2004MNRAS.350..517S}, there are two classifications of voids, i.e., big \emph{void-in-void} voids embedded in larger underdense regions (larger distinct voids) and small \emph{void-in-cloud} voids embedded within a larger-scale overdensity. Here, \emph{void-in-void} voids form at early epoch and then collide and merge with one another at late epoch, forming a larger distinct void. In this process, matter between them is squeezed and evacuated along walls and filaments towards the enclosing boundary of the larger newly formed void, leaving faint and gradually fading imprint of the initial internal substructures. And the same basic process repeats as this rearrangement of structure develops to a larger scale. While, \emph{void-in-cloud} voids are squeezed by larger-scale overdensity and will vanish when the region around them has collapsed completely.

Specifically, we find $\Delta V_3 < 0$ for $\rho/\overline{\rho} \lesssim 0.05$ due to the decline in number of isolated underdense troughs (bubbles), corresponding to the suppression effect on number function of big voids in neutrino cosmology \cite{2015JCAP...11..018M, 2019MNRAS.488.4413K}. The cosmic neutrinos slow down the \emph{void-in-void} process, making the faint regions in sheet-like structures denser and evener. As a result, with $\rho/\overline {\rho}$ adjusted to higher value, it gets harder to pierce through their thinner parts to form tunnels in the excursion set's surface. At $\rho/\overline{\rho} \approx 0.5$, we find that $V_3$ stop rising and start falling, which can be reasonably attributed to the emergence of tunnels. Therefore, in the range of $0.05 \lesssim \rho/\overline{\rho} \lesssim 0.2$, we see $\Delta V_3 < 0$ due to the decrease in number of tunnels in $\nu\Lambda$CDM cosmology.

Neutrino suppress matter clustering on small scale, which has been well understood by a minimum of "spoon" shape around $k = 1$ $h$Mpc$^{-1}$ (corresponding to the size of massive halos) on $P_{m}^{\nu}/P_{m}^{fiducial}$, at $z \sim 0$ (e.g., \cite{2011MNRAS.410.1647A, 2014JCAP...12..053M}). Due to this neutrino effect, matter in virialized objects is smeared around and filled in the \emph{void-in-cloud} voids. In addition, this smeared matter also patch the relatively thin parts in the denser sheet-like structures. Therefore, with $\rho/\overline{\rho}$ rising ($\rho/\overline{\rho} > 0.2$), we first see a similar trend as we see in the former two scenarios but with mild amplitude; when $\rho/\overline{\rho}$ being high enough, the excursion set will turn into isolated virialized density peaks (balls), finally we see $V_3$ goes to below zero, corresponding to the suppression effect on mass function of massive halos in neutrino cosmology \cite{2012PhRvD..85f3521I, 2013JCAP...12..012C, 2018JCAP...03..049L} ; i.e., $\Delta V_3 < 0$, then $\Delta V_3 > 0$ and finally $\Delta V_3 < 0$. With $\rho/\overline{\rho}$ rising higher and higher, we find $\Delta V_3$ approximates to zero asymptotically, since the small halos with higher concentrations \cite{1997ApJ...490..493N} are less impacted by massive neutrinos, corresponding to the upturn at high $k$ ($> 1$ $h$Mpc$^{-1}$) on $P_{m}^{\nu}/P_{m}^{fiducial}$ \cite{2018JCAP...03..049L}.

\begin{figure*}[ht]
\hspace*{-0.7cm}
{\includegraphics[angle=0,scale=1.275,trim=0cm 0cm 0cm 0cm,clip=true]{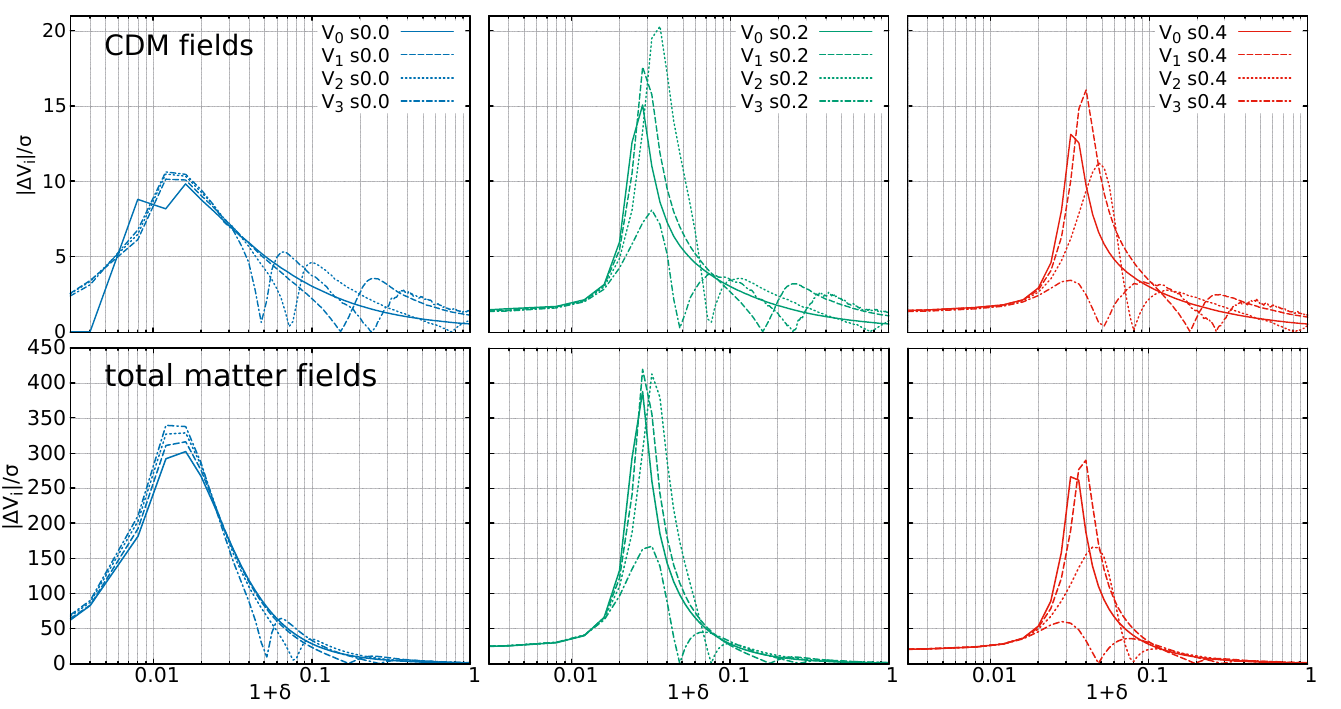}}
\caption{The signal-to-noise ratios of neutrino signatures ($\Sigma m_{\nu} = 0.05$ eV) as functions of $1 + \delta$ (i.e., $\rho/\overline{\rho}$), at $z = 0.01$. The top and bottom panels are for CDM and total matter fields, respectively. Different columns are for different smoothing scales, i.e., $0L_g$, 0.2$L_g$, 0.4$L_g$.
}
\label{fig:mf2}
\end{figure*}

\section{Smoothing effects}

Gaussian smoothing is usually used to reduce noise contribution to the fields for MFs measurements (e.g., \cite{Schmalzing_1997, 2013MNRAS.429.2104D, 2017PhRvL.118r1301F}). While, this process also erase non-Gaussian information from original fields \cite{2013MNRAS.429.2104D}, which can downgrade the discriminative power of MFs when resolving neutrino signatures. In our work, we note that appropriately smoothing density fields (e.g., $R_g = 0.2L_g$) can improve the signal-to-noise (S/N) ratios of neutrino signals, $|\Delta V_i|/\sigma$, on $\Delta V_0$, $\Delta V_1$ and $\Delta V_2$ within a narrow range around $\rho/\overline{\rho} = 0.03$, while it otherwise depresses the S/N ratios in other ranges. Conversely, it seems to definitely decrease the S/N ratios on $\Delta V_3$ in the whole ranges, regardless of $R_g$. This may be due to that topology ($V_3$) is more susceptible to this artificial smearing of LSS than other $V_i$s, on the premise of noise reduction. From the decline ratios of $\Delta V_i$s' amplitudes and the S/N ratios of neutrino signatures on $\Delta V_i$s caused by different smoothing in Figure~\ref{fig:mf1} and  Figure~\ref{fig:mf2}, we can preliminary infer that the sensitivities of MFs to nonGaussianity (and to $\sum m_{\nu}$) roughly obey $V_1 < V_2 < V_3 \lesssim V_4$, which is consistent with previous studies on $2D$ MFs of weak lensing and CMB (e.g., \cite{2012PhRvD..85j3513K, 2013MNRAS.429.2104D, 2019JCAP...06..019M}).

\section{Summary and Conclusion}
In the past decade, cosmology has achieved great success in neutrino mass constraint. However, further improvement of neutrino mass constraint using LSS is mainly hindered by the challenges of statistical methods and systematics. The key problems are as follows:

\begin{enumerate}[label=(\roman*)]
	\item LSS has evolved to be highly non-Gaussian in the later Universe. Traditional methods for extracting neutrino information are based on two-point statistics. These traditional methods only can probe Gaussian information from LSS, missing substantial higher-order information.
	\item Moreover, the neutrino signals extracted by traditional methods are mainly contributed by the neutrino effects on the small scales of high-density regions of LSS. However, this neutrino information suffers from the contaminations by tricky nonlinear effects and baryonic physics effects, etc.
\end{enumerate}

Toward solving these critical problems (for improving constraining power on $\sum m_{\nu}$ in data analysis of up-coming LSS surveys), we propose an alternative powerful non-Gaussian probe of neutrino effects on LSS, i.e., Minkowski functionals (MFs), in this work. This tool not only has strong statistical power, but also has strong robustness to systematics. It can extract full information encoded in LSS, circumventing a more complicated N-point statistics formalism. Better yet, the neutrino information extracted by this method is mainly from low-density regions \cite{2013MNRAS.431.3670V} (because these regions with the highest neutrino to CDM density ratios should be more sensitive to neutrinos \cite{2020arXiv200111024M}), which potentially makes the extracted neutrino signals well avoid various contaminations of high-density regions. Therefore, the problems faced in the past are expected to be greatly alleviated.

By using this novel method, for the first time, we comprehensively studied subtle neutrino effects on the morphology of LSS, which further deepens our understanding of neutrino effects and provides essential and critical information for accurate modeling of neutrino effects in the future.  For an ideal LSS survey of volume $\sim1.73$ Gpc$^3$/$h^3$, we show a compelling result that the neutrino signals can be extracted with a significant level up to $\thicksim 10\sigma$ and $\thicksim 300\sigma$ for CDM and total matter density fields, respectively, with an individual MF measurement (cf. Figure~\ref{fig:mf2}). These results demonstrate its great potential for much improving neutrino mass constraint in the data analysis of forth-coming LSS surveys.

 Nevertheless, we have to mention that matter fields cannot be directly obtained from galaxy surveys. In reality, underlying matter fields are mapped by biased tracers, i.e., halos and galaxies. Here, our results can be treated as the theoretical upper limit of neutrino effects on halo/galaxy distribution. In view of the strong statistical power of MFs \cite{2017PhRvL.118r1301F}, these neutrino probings can probably survive in ambitious galaxy surveys with large galaxy number densities (e.g., BGS sample in DESI \cite{2016arXiv161100036D}). We postpone such a comprehensive study in an ongoing work, where stochasticity is reduced in the mass-weighted halo field \cite{2010PhRvD..82d3515H} and mock galaxies are constructed by halo occupation distribution technique \cite{2005ApJ...633..791Z}.

\section{Acknowledgements}
We thank the anonymous referee for the useful comments and suggestions. Y.L. thanks Thomas
Buchert and Wenjuan Fang for helpful communications. This work was supported by the National
Key Basic Research and Development Program of China (No. 2018YFA0404504), and the National
Science Foundation of China (grants No. 11773048, 11621303, 11890691). H.R.Y. is supported by
National Science Foundation of China 11903021.

\bibliography{mf.bib}

\begin{thebibliography}{81}%
\makeatletter
\providecommand \@ifxundefined [1]{%
 \@ifx{#1\undefined}
}%
\providecommand \@ifnum [1]{%
 \ifnum #1\expandafter \@firstoftwo
 \else \expandafter \@secondoftwo
 \fi
}%
\providecommand \@ifx [1]{%
 \ifx #1\expandafter \@firstoftwo
 \else \expandafter \@secondoftwo
 \fi
}%
\providecommand \natexlab [1]{#1}%
\providecommand \enquote  [1]{``#1''}%
\providecommand \bibnamefont  [1]{#1}%
\providecommand \bibfnamefont [1]{#1}%
\providecommand \citenamefont [1]{#1}%
\providecommand \href@noop [0]{\@secondoftwo}%
\providecommand \href [0]{\begingroup \@sanitize@url \@href}%
\providecommand \@href[1]{\@@startlink{#1}\@@href}%
\providecommand \@@href[1]{\endgroup#1\@@endlink}%
\providecommand \@sanitize@url [0]{\catcode `\\12\catcode `\$12\catcode
  `\&12\catcode `\#12\catcode `\^12\catcode `\_12\catcode `\%12\relax}%
\providecommand \@@startlink[1]{}%
\providecommand \@@endlink[0]{}%
\providecommand \url  [0]{\begingroup\@sanitize@url \@url }%
\providecommand \@url [1]{\endgroup\@href {#1}{\urlprefix }}%
\providecommand \urlprefix  [0]{URL }%
\providecommand \Eprint [0]{\href }%
\providecommand \doibase [0]{http://dx.doi.org/}%
\providecommand \selectlanguage [0]{\@gobble}%
\providecommand \bibinfo  [0]{\@secondoftwo}%
\providecommand \bibfield  [0]{\@secondoftwo}%
\providecommand \translation [1]{[#1]}%
\providecommand \BibitemOpen [0]{}%
\providecommand \bibitemStop [0]{}%
\providecommand \bibitemNoStop [0]{.\EOS\space}%
\providecommand \EOS [0]{\spacefactor3000\relax}%
\providecommand \BibitemShut  [1]{\csname bibitem#1\endcsname}%
\let\auto@bib@innerbib\@empty
\bibitem [{200(2006)}]{2006PhR...427..257A}%
  \BibitemOpen
  \href {\doibase 10.1016/j.physrep.2005.12.006} {\bibfield  {journal}
  {\bibinfo  {journal} {Physics Reports}\ }\textbf {\bibinfo {volume} {427}},\
  \bibinfo {pages} {257} (\bibinfo {year} {2006})}\BibitemShut {NoStop}%
\bibitem [{\citenamefont {{Becker-Szendy}}\ \emph {et~al.}(1992)\citenamefont
  {{Becker-Szendy}}, \citenamefont {{Bratton}},\ and\ \citenamefont {{Casper et
  al.}}}]{1992PhRvD..46.3720B}%
  \BibitemOpen
  \bibfield  {author} {\bibinfo {author} {\bibfnamefont {R.}~\bibnamefont
  {{Becker-Szendy}}}, \bibinfo {author} {\bibfnamefont {C.~B.}\ \bibnamefont
  {{Bratton}}}, \ and\ \bibinfo {author} {\bibfnamefont {D.}~\bibnamefont
  {{Casper et al.}}},\ }\href {\doibase 10.1103/PhysRevD.46.3720} {\bibfield
  {journal} {\bibinfo  {journal} {Physical Review D}\ }\textbf {\bibinfo
  {volume} {46}},\ \bibinfo {pages} {3720} (\bibinfo {year}
  {1992})}\BibitemShut {NoStop}%
\bibitem [{\citenamefont {{Fukuda}}\ \emph {et~al.}(1998)\citenamefont
  {{Fukuda}}, \citenamefont {{Hayakawa}}, \citenamefont {{Ichihara}},\ and\
  \citenamefont {{Inoue et al.}}}]{1998PhRvL..81.1562F}%
  \BibitemOpen
  \bibfield  {author} {\bibinfo {author} {\bibfnamefont {Y.}~\bibnamefont
  {{Fukuda}}}, \bibinfo {author} {\bibfnamefont {T.}~\bibnamefont
  {{Hayakawa}}}, \bibinfo {author} {\bibfnamefont {E.}~\bibnamefont
  {{Ichihara}}}, \ and\ \bibinfo {author} {\bibfnamefont {K.}~\bibnamefont
  {{Inoue et al.}}},\ }\href {\doibase 10.1103/PhysRevLett.81.1562} {\bibfield
  {journal} {\bibinfo  {journal} {Physical Review Letters}\ }\textbf {\bibinfo
  {volume} {81}},\ \bibinfo {pages} {1562} (\bibinfo {year} {1998})},\ \Eprint
  {http://arxiv.org/abs/hep-ex/9807003} {arXiv:hep-ex/9807003 [hep-ex]}
  \BibitemShut {NoStop}%
\bibitem [{\citenamefont {{Ahmed}}\ \emph {et~al.}(2004)\citenamefont
  {{Ahmed}}, \citenamefont {{Anthony}},\ and\ \citenamefont {{Beier et
  al.}}}]{2004PhRvL..92r1301A}%
  \BibitemOpen
  \bibfield  {author} {\bibinfo {author} {\bibfnamefont {S.~N.}\ \bibnamefont
  {{Ahmed}}}, \bibinfo {author} {\bibfnamefont {A.~E.}\ \bibnamefont
  {{Anthony}}}, \ and\ \bibinfo {author} {\bibfnamefont {E.~W.}\ \bibnamefont
  {{Beier et al.}}},\ }\href {\doibase 10.1103/PhysRevLett.92.181301}
  {\bibfield  {journal} {\bibinfo  {journal} {Physical Review Letters}\
  }\textbf {\bibinfo {volume} {92}},\ \bibinfo {eid} {181301} (\bibinfo {year}
  {2004})},\ \Eprint {http://arxiv.org/abs/nucl-ex/0309004}
  {arXiv:nucl-ex/0309004 [nucl-ex]} \BibitemShut {NoStop}%
\bibitem [{\citenamefont {{Adamson}}\ \emph {et~al.}(2008)\citenamefont
  {{Adamson}}, \citenamefont {{Andreopoulos}},\ and\ \citenamefont {{Arms et
  al.}}}]{2008PhRvL.101m1802A}%
  \BibitemOpen
  \bibfield  {author} {\bibinfo {author} {\bibfnamefont {P.}~\bibnamefont
  {{Adamson}}}, \bibinfo {author} {\bibfnamefont {C.}~\bibnamefont
  {{Andreopoulos}}}, \ and\ \bibinfo {author} {\bibfnamefont {K.~E.}\
  \bibnamefont {{Arms et al.}}},\ }\href {\doibase
  10.1103/PhysRevLett.101.131802} {\bibfield  {journal} {\bibinfo  {journal}
  {Physical Review Letters}\ }\textbf {\bibinfo {volume} {101}},\ \bibinfo
  {eid} {131802} (\bibinfo {year} {2008})},\ \Eprint
  {http://arxiv.org/abs/0806.2237} {arXiv:0806.2237 [hep-ex]} \BibitemShut
  {NoStop}%
\bibitem [{\citenamefont {{Wolf}}\ and\ \citenamefont {{Katrin
  Collaboration}}(2010)}]{2010NIMPA.623..442W}%
  \BibitemOpen
  \bibfield  {author} {\bibinfo {author} {\bibfnamefont {J.}~\bibnamefont
  {{Wolf}}}\ and\ \bibinfo {author} {\bibnamefont {{Katrin Collaboration}}},\
  }\href {\doibase 10.1016/j.nima.2010.03.030} {\bibfield  {journal} {\bibinfo
  {journal} {Nuclear Instruments and Methods in Physics Research A}\ }\textbf
  {\bibinfo {volume} {623}},\ \bibinfo {pages} {442} (\bibinfo {year}
  {2010})},\ \Eprint {http://arxiv.org/abs/0810.3281} {arXiv:0810.3281
  [physics.ins-det]} \BibitemShut {NoStop}%
\bibitem [{\citenamefont {{Ashtari Esfahani}}\ and\ \citenamefont {{Asner et
  al.}}(2017)}]{2017JPhG...44e4004A}%
  \BibitemOpen
  \bibfield  {author} {\bibinfo {author} {\bibfnamefont {A.}~\bibnamefont
  {{Ashtari Esfahani}}}\ and\ \bibinfo {author} {\bibfnamefont {D.~M.}\
  \bibnamefont {{Asner et al.}}},\ }\href {\doibase 10.1088/1361-6471/aa5b4f}
  {\bibfield  {journal} {\bibinfo  {journal} {Journal of Physics G Nuclear
  Physics}\ }\textbf {\bibinfo {volume} {44}},\ \bibinfo {eid} {054004}
  (\bibinfo {year} {2017})},\ \Eprint {http://arxiv.org/abs/1703.02037}
  {arXiv:1703.02037 [physics.ins-det]} \BibitemShut {NoStop}%
\bibitem [{\citenamefont {{Planck Collaboration}}\ and\ \citenamefont {{Aghanim
  et al.}}(2018)}]{2018arXiv180706209P}%
  \BibitemOpen
  \bibfield  {author} {\bibinfo {author} {\bibnamefont {{Planck
  Collaboration}}}\ and\ \bibinfo {author} {\bibfnamefont {N.}~\bibnamefont
  {{Aghanim et al.}}},\ }\href@noop {} {\bibfield  {journal} {\bibinfo
  {journal} {arXiv e-prints}\ ,\ \bibinfo {eid} {arXiv:1807.06209}} (\bibinfo
  {year} {2018})},\ \Eprint {http://arxiv.org/abs/1807.06209} {arXiv:1807.06209
  [astro-ph.CO]} \BibitemShut {NoStop}%
\bibitem [{\citenamefont
  {\href{https://www.skatelescope.org}{https://www.skatelescope.org}}()}]{SKA}%
  \BibitemOpen
  \bibfield  {author} {\bibinfo {author} {\bibnamefont
  {\href{https://www.skatelescope.org}{https://www.skatelescope.org}}},\
  }\href@noop {} {\ }\BibitemShut {NoStop}%
\bibitem [{\citenamefont {{DESI Collaboration}}\ and\ \citenamefont {{Aghamousa
  et al.}}(2016)}]{2016arXiv161100036D}%
  \BibitemOpen
  \bibfield  {author} {\bibinfo {author} {\bibnamefont {{DESI Collaboration}}}\
  and\ \bibinfo {author} {\bibfnamefont {A.}~\bibnamefont {{Aghamousa et
  al.}}},\ }\href@noop {} {\bibfield  {journal} {\bibinfo  {journal} {arXiv
  e-prints}\ ,\ \bibinfo {eid} {arXiv:1611.00036}} (\bibinfo {year} {2016})},\
  \Eprint {http://arxiv.org/abs/1611.00036} {arXiv:1611.00036 [astro-ph.IM]}
  \BibitemShut {NoStop}%
\bibitem [{\citenamefont {{LSST Science Collaboration}}\ and\ \citenamefont
  {{Abell et al.}}(2009)}]{2009arXiv0912.0201L}%
  \BibitemOpen
  \bibfield  {author} {\bibinfo {author} {\bibnamefont {{LSST Science
  Collaboration}}}\ and\ \bibinfo {author} {\bibfnamefont {P.~A.}\ \bibnamefont
  {{Abell et al.}}},\ }\href@noop {} {\bibfield  {journal} {\bibinfo  {journal}
  {arXiv e-prints}\ ,\ \bibinfo {eid} {arXiv:0912.0201}} (\bibinfo {year}
  {2009})},\ \Eprint {http://arxiv.org/abs/0912.0201} {arXiv:0912.0201
  [astro-ph.IM]} \BibitemShut {NoStop}%
\bibitem [{\citenamefont
  {\href{http://wfirst.gsfc.nasa.gov}{http://wfirst.gsfc.nasa.gov}}()}]{wfirst}%
  \BibitemOpen
  \bibfield  {author} {\bibinfo {author} {\bibnamefont
  {\href{http://wfirst.gsfc.nasa.gov}{http://wfirst.gsfc.nasa.gov}}},\
  }\href@noop {} {\ }\BibitemShut {NoStop}%
\bibitem [{\citenamefont
  {\href{https://www.euclid-ec.org}{https://www.euclid-ec.org}}()}]{euclid}%
  \BibitemOpen
  \bibfield  {author} {\bibinfo {author} {\bibnamefont
  {\href{https://www.euclid-ec.org}{https://www.euclid-ec.org}}},\ }\href@noop
  {} {\ }\BibitemShut {NoStop}%
\bibitem [{\citenamefont {{Ade}}\ and\ \citenamefont {{Aguirre et
  al.}}(2019)}]{2019JCAP...02..056A}%
  \BibitemOpen
  \bibfield  {author} {\bibinfo {author} {\bibfnamefont {P.}~\bibnamefont
  {{Ade}}}\ and\ \bibinfo {author} {\bibfnamefont {J.}~\bibnamefont {{Aguirre
  et al.}}},\ }\href {\doibase 10.1088/1475-7516/2019/02/056} {\bibfield
  {journal} {\bibinfo  {journal} {\jcap}\ }\textbf {\bibinfo {volume} {2019}},\
  \bibinfo {eid} {056} (\bibinfo {year} {2019})},\ \Eprint
  {http://arxiv.org/abs/1808.07445} {arXiv:1808.07445 [astro-ph.CO]}
  \BibitemShut {NoStop}%
\bibitem [{\citenamefont {{Abazajian}}\ \emph {et~al.}(2016)\citenamefont
  {{Abazajian}}, \citenamefont {{Adshead}},\ and\ \citenamefont {{Ahmed et
  al.}}}]{2016arXiv161002743A}%
  \BibitemOpen
  \bibfield  {author} {\bibinfo {author} {\bibfnamefont {K.~N.}\ \bibnamefont
  {{Abazajian}}}, \bibinfo {author} {\bibfnamefont {P.}~\bibnamefont
  {{Adshead}}}, \ and\ \bibinfo {author} {\bibfnamefont {Z.}~\bibnamefont
  {{Ahmed et al.}}},\ }\href@noop {} {\bibfield  {journal} {\bibinfo  {journal}
  {arXiv e-prints}\ ,\ \bibinfo {eid} {arXiv:1610.02743}} (\bibinfo {year}
  {2016})},\ \Eprint {http://arxiv.org/abs/1610.02743} {arXiv:1610.02743
  [astro-ph.CO]} \BibitemShut {NoStop}%
\bibitem [{\citenamefont {{Hu}}\ \emph {et~al.}(1998)\citenamefont {{Hu}},
  \citenamefont {{Eisenstein}},\ and\ \citenamefont
  {{Tegmark}}}]{1998PhRvL..80.5255H}%
  \BibitemOpen
  \bibfield  {author} {\bibinfo {author} {\bibfnamefont {W.}~\bibnamefont
  {{Hu}}}, \bibinfo {author} {\bibfnamefont {D.~J.}\ \bibnamefont
  {{Eisenstein}}}, \ and\ \bibinfo {author} {\bibfnamefont {M.}~\bibnamefont
  {{Tegmark}}},\ }\href {\doibase 10.1103/PhysRevLett.80.5255} {\bibfield
  {journal} {\bibinfo  {journal} {\prl}\ }\textbf {\bibinfo {volume} {80}},\
  \bibinfo {pages} {5255} (\bibinfo {year} {1998})},\ \Eprint
  {http://arxiv.org/abs/astro-ph/9712057} {arXiv:astro-ph/9712057 [astro-ph]}
  \BibitemShut {NoStop}%
\bibitem [{\citenamefont {Lesgourgues}\ and\ \citenamefont
  {Pastor}(2006)}]{article}%
  \BibitemOpen
  \bibfield  {author} {\bibinfo {author} {\bibfnamefont {J.}~\bibnamefont
  {Lesgourgues}}\ and\ \bibinfo {author} {\bibfnamefont {S.}~\bibnamefont
  {Pastor}},\ }\href {\doibase 10.1016/j.physrep.2006.04.001} {\bibfield
  {journal} {\bibinfo  {journal} {Physics Reports}\ }\textbf {\bibinfo {volume}
  {429}} (\bibinfo {year} {2006}),\ 10.1016/j.physrep.2006.04.001}\BibitemShut
  {NoStop}%
\bibitem [{\citenamefont {{Wong}}(2011)}]{2011ARNPS..61...69W}%
  \BibitemOpen
  \bibfield  {author} {\bibinfo {author} {\bibfnamefont {Y.~Y.~Y.}\
  \bibnamefont {{Wong}}},\ }\href {\doibase 10.1146/annurev-nucl-102010-130252}
  {\bibfield  {journal} {\bibinfo  {journal} {Annual Review of Nuclear and
  Particle Science}\ }\textbf {\bibinfo {volume} {61}},\ \bibinfo {pages} {69}
  (\bibinfo {year} {2011})},\ \Eprint {http://arxiv.org/abs/1111.1436}
  {arXiv:1111.1436 [astro-ph.CO]} \BibitemShut {NoStop}%
\bibitem [{\citenamefont {{Lesgourgues}}\ \emph {et~al.}(2013)\citenamefont
  {{Lesgourgues}}, \citenamefont {{Mangano}}, \citenamefont {{Miele}},\ and\
  \citenamefont {{Pastor}}}]{2013neco.book.....L}%
  \BibitemOpen
  \bibfield  {author} {\bibinfo {author} {\bibfnamefont {J.}~\bibnamefont
  {{Lesgourgues}}}, \bibinfo {author} {\bibfnamefont {G.}~\bibnamefont
  {{Mangano}}}, \bibinfo {author} {\bibfnamefont {G.}~\bibnamefont {{Miele}}},
  \ and\ \bibinfo {author} {\bibfnamefont {S.}~\bibnamefont {{Pastor}}},\
  }\href@noop {} {\emph {\bibinfo {title} {{Neutrino Cosmology}}}}\ (\bibinfo
  {year} {2013})\BibitemShut {NoStop}%
\bibitem [{\citenamefont {{Saito}}\ \emph {et~al.}(2008)\citenamefont
  {{Saito}}, \citenamefont {{Takada}},\ and\ \citenamefont
  {{Taruya}}}]{2008PhRvL.100s1301S}%
  \BibitemOpen
  \bibfield  {author} {\bibinfo {author} {\bibfnamefont {S.}~\bibnamefont
  {{Saito}}}, \bibinfo {author} {\bibfnamefont {M.}~\bibnamefont {{Takada}}}, \
  and\ \bibinfo {author} {\bibfnamefont {A.}~\bibnamefont {{Taruya}}},\ }\href
  {\doibase 10.1103/PhysRevLett.100.191301} {\bibfield  {journal} {\bibinfo
  {journal} {\prl}\ }\textbf {\bibinfo {volume} {100}},\ \bibinfo {eid}
  {191301} (\bibinfo {year} {2008})},\ \Eprint {http://arxiv.org/abs/0801.0607}
  {arXiv:0801.0607 [astro-ph]} \BibitemShut {NoStop}%
\bibitem [{\citenamefont {{Petracca}}\ \emph {et~al.}(2016)\citenamefont
  {{Petracca}}, \citenamefont {{Marulli}}, \citenamefont {{Moscardini}},
  \citenamefont {{Cimatti}}, \citenamefont {{Carbone}},\ and\ \citenamefont
  {{Angulo}}}]{2016MNRAS.462.4208P}%
  \BibitemOpen
  \bibfield  {author} {\bibinfo {author} {\bibfnamefont {F.}~\bibnamefont
  {{Petracca}}}, \bibinfo {author} {\bibfnamefont {F.}~\bibnamefont
  {{Marulli}}}, \bibinfo {author} {\bibfnamefont {L.}~\bibnamefont
  {{Moscardini}}}, \bibinfo {author} {\bibfnamefont {A.}~\bibnamefont
  {{Cimatti}}}, \bibinfo {author} {\bibfnamefont {C.}~\bibnamefont
  {{Carbone}}}, \ and\ \bibinfo {author} {\bibfnamefont {R.~E.}\ \bibnamefont
  {{Angulo}}},\ }\href {\doibase 10.1093/mnras/stw1948} {\bibfield  {journal}
  {\bibinfo  {journal} {\mnras}\ }\textbf {\bibinfo {volume} {462}},\ \bibinfo
  {pages} {4208} (\bibinfo {year} {2016})},\ \Eprint
  {http://arxiv.org/abs/1512.06139} {arXiv:1512.06139 [astro-ph.CO]}
  \BibitemShut {NoStop}%
\bibitem [{\citenamefont {{Archidiacono}}\ \emph {et~al.}(2017)\citenamefont
  {{Archidiacono}}, \citenamefont {{Brinckmann}}, \citenamefont
  {{Lesgourgues}},\ and\ \citenamefont {{Poulin}}}]{2017JCAP...02..052A}%
  \BibitemOpen
  \bibfield  {author} {\bibinfo {author} {\bibfnamefont {M.}~\bibnamefont
  {{Archidiacono}}}, \bibinfo {author} {\bibfnamefont {T.}~\bibnamefont
  {{Brinckmann}}}, \bibinfo {author} {\bibfnamefont {J.}~\bibnamefont
  {{Lesgourgues}}}, \ and\ \bibinfo {author} {\bibfnamefont {V.}~\bibnamefont
  {{Poulin}}},\ }\href {\doibase 10.1088/1475-7516/2017/02/052} {\bibfield
  {journal} {\bibinfo  {journal} {\jcap}\ }\textbf {\bibinfo {volume} {2017}},\
  \bibinfo {eid} {052} (\bibinfo {year} {2017})},\ \Eprint
  {http://arxiv.org/abs/1610.09852} {arXiv:1610.09852 [astro-ph.CO]}
  \BibitemShut {NoStop}%
\bibitem [{\citenamefont {{Chudaykin}}\ and\ \citenamefont
  {{Ivanov}}(2019)}]{2019arXiv190706666C}%
  \BibitemOpen
  \bibfield  {author} {\bibinfo {author} {\bibfnamefont {A.}~\bibnamefont
  {{Chudaykin}}}\ and\ \bibinfo {author} {\bibfnamefont {M.~M.}\ \bibnamefont
  {{Ivanov}}},\ }\href@noop {} {\bibfield  {journal} {\bibinfo  {journal}
  {arXiv e-prints}\ ,\ \bibinfo {eid} {arXiv:1907.06666}} (\bibinfo {year}
  {2019})},\ \Eprint {http://arxiv.org/abs/1907.06666} {arXiv:1907.06666
  [astro-ph.CO]} \BibitemShut {NoStop}%
\bibitem [{\citenamefont {{Banerjee}}\ \emph {et~al.}(2019)\citenamefont
  {{Banerjee}}, \citenamefont {{Castorina}},\ and\ \citenamefont
  {{Villaescusa-Navarro et al.}}}]{2019arXiv190706598B}%
  \BibitemOpen
  \bibfield  {author} {\bibinfo {author} {\bibfnamefont {A.}~\bibnamefont
  {{Banerjee}}}, \bibinfo {author} {\bibfnamefont {E.}~\bibnamefont
  {{Castorina}}}, \ and\ \bibinfo {author} {\bibfnamefont {F.}~\bibnamefont
  {{Villaescusa-Navarro et al.}}},\ }\href@noop {} {\bibfield  {journal}
  {\bibinfo  {journal} {arXiv e-prints}\ ,\ \bibinfo {eid} {arXiv:1907.06598}}
  (\bibinfo {year} {2019})},\ \Eprint {http://arxiv.org/abs/1907.06598}
  {arXiv:1907.06598 [astro-ph.CO]} \BibitemShut {NoStop}%
\bibitem [{\citenamefont {{Parimbelli}}\ \emph {et~al.}(2019)\citenamefont
  {{Parimbelli}}, \citenamefont {{Viel}},\ and\ \citenamefont
  {{Sefusatti}}}]{2019JCAP...01..010P}%
  \BibitemOpen
  \bibfield  {author} {\bibinfo {author} {\bibfnamefont {G.}~\bibnamefont
  {{Parimbelli}}}, \bibinfo {author} {\bibfnamefont {M.}~\bibnamefont
  {{Viel}}}, \ and\ \bibinfo {author} {\bibfnamefont {E.}~\bibnamefont
  {{Sefusatti}}},\ }\href {\doibase 10.1088/1475-7516/2019/01/010} {\bibfield
  {journal} {\bibinfo  {journal} {\jcap}\ }\textbf {\bibinfo {volume} {2019}},\
  \bibinfo {eid} {010} (\bibinfo {year} {2019})},\ \Eprint
  {http://arxiv.org/abs/1809.06634} {arXiv:1809.06634 [astro-ph.CO]}
  \BibitemShut {NoStop}%
\bibitem [{\citenamefont {{Villaescusa-Navarro}}\ \emph
  {et~al.}(2018)\citenamefont {{Villaescusa-Navarro}}, \citenamefont
  {{Banerjee}},\ and\ \citenamefont {{Dalal et al.}}}]{2018ApJ...861...53V}%
  \BibitemOpen
  \bibfield  {author} {\bibinfo {author} {\bibfnamefont {F.}~\bibnamefont
  {{Villaescusa-Navarro}}}, \bibinfo {author} {\bibfnamefont {A.}~\bibnamefont
  {{Banerjee}}}, \ and\ \bibinfo {author} {\bibfnamefont {N.}~\bibnamefont
  {{Dalal et al.}}},\ }\href {\doibase 10.3847/1538-4357/aac6bf} {\bibfield
  {journal} {\bibinfo  {journal} {\apj}\ }\textbf {\bibinfo {volume} {861}},\
  \bibinfo {eid} {53} (\bibinfo {year} {2018})},\ \Eprint
  {http://arxiv.org/abs/1708.01154} {arXiv:1708.01154 [astro-ph.CO]}
  \BibitemShut {NoStop}%
\bibitem [{\citenamefont {{Fang}}\ \emph {et~al.}(2017)\citenamefont {{Fang}},
  \citenamefont {{Li}},\ and\ \citenamefont {{Zhao}}}]{2017PhRvL.118r1301F}%
  \BibitemOpen
  \bibfield  {author} {\bibinfo {author} {\bibfnamefont {W.}~\bibnamefont
  {{Fang}}}, \bibinfo {author} {\bibfnamefont {B.}~\bibnamefont {{Li}}}, \ and\
  \bibinfo {author} {\bibfnamefont {G.-B.}\ \bibnamefont {{Zhao}}},\ }\href
  {\doibase 10.1103/PhysRevLett.118.181301} {\bibfield  {journal} {\bibinfo
  {journal} {\prl}\ }\textbf {\bibinfo {volume} {118}},\ \bibinfo {eid}
  {181301} (\bibinfo {year} {2017})},\ \Eprint
  {http://arxiv.org/abs/1704.02325} {arXiv:1704.02325 [astro-ph.CO]}
  \BibitemShut {NoStop}%
\bibitem [{\citenamefont {{Hagstotz}}\ \emph {et~al.}(2019)\citenamefont
  {{Hagstotz}}, \citenamefont {{Gronke}}, \citenamefont {{Mota}},\ and\
  \citenamefont {{Baldi}}}]{2019A&A...629A..46H}%
  \BibitemOpen
  \bibfield  {author} {\bibinfo {author} {\bibfnamefont {S.}~\bibnamefont
  {{Hagstotz}}}, \bibinfo {author} {\bibfnamefont {M.}~\bibnamefont
  {{Gronke}}}, \bibinfo {author} {\bibfnamefont {D.~F.}\ \bibnamefont
  {{Mota}}}, \ and\ \bibinfo {author} {\bibfnamefont {M.}~\bibnamefont
  {{Baldi}}},\ }\href {\doibase 10.1051/0004-6361/201935213} {\bibfield
  {journal} {\bibinfo  {journal} {\aap}\ }\textbf {\bibinfo {volume} {629}},\
  \bibinfo {eid} {A46} (\bibinfo {year} {2019})},\ \Eprint
  {http://arxiv.org/abs/1902.01868} {arXiv:1902.01868 [astro-ph.CO]}
  \BibitemShut {NoStop}%
\bibitem [{\citenamefont {{Wright}}\ \emph {et~al.}(2019)\citenamefont
  {{Wright}}, \citenamefont {{Koyama}}, \citenamefont {{Winther}},\ and\
  \citenamefont {{Zhao}}}]{2019JCAP...06..040W}%
  \BibitemOpen
  \bibfield  {author} {\bibinfo {author} {\bibfnamefont {B.~S.}\ \bibnamefont
  {{Wright}}}, \bibinfo {author} {\bibfnamefont {K.}~\bibnamefont {{Koyama}}},
  \bibinfo {author} {\bibfnamefont {H.~A.}\ \bibnamefont {{Winther}}}, \ and\
  \bibinfo {author} {\bibfnamefont {G.-B.}\ \bibnamefont {{Zhao}}},\ }\href
  {\doibase 10.1088/1475-7516/2019/06/040} {\bibfield  {journal} {\bibinfo
  {journal} {\jcap}\ }\textbf {\bibinfo {volume} {2019}},\ \bibinfo {eid} {040}
  (\bibinfo {year} {2019})},\ \Eprint {http://arxiv.org/abs/1902.10692}
  {arXiv:1902.10692 [astro-ph.CO]} \BibitemShut {NoStop}%
\bibitem [{\citenamefont {{Zhu}}\ and\ \citenamefont
  {{Castorina}}(2019)}]{2019arXiv190500361Z}%
  \BibitemOpen
  \bibfield  {author} {\bibinfo {author} {\bibfnamefont {H.-M.}\ \bibnamefont
  {{Zhu}}}\ and\ \bibinfo {author} {\bibfnamefont {E.}~\bibnamefont
  {{Castorina}}},\ }\href@noop {} {\bibfield  {journal} {\bibinfo  {journal}
  {arXiv e-prints}\ ,\ \bibinfo {eid} {arXiv:1905.00361}} (\bibinfo {year}
  {2019})},\ \Eprint {http://arxiv.org/abs/1905.00361} {arXiv:1905.00361
  [astro-ph.CO]} \BibitemShut {NoStop}%
\bibitem [{\citenamefont {{Yu}}\ \emph {et~al.}(2019)\citenamefont {{Yu}},
  \citenamefont {{Pen}},\ and\ \citenamefont {{Wang}}}]{2019PhRvD..99l3532Y}%
  \BibitemOpen
  \bibfield  {author} {\bibinfo {author} {\bibfnamefont {H.-R.}\ \bibnamefont
  {{Yu}}}, \bibinfo {author} {\bibfnamefont {U.-L.}\ \bibnamefont {{Pen}}}, \
  and\ \bibinfo {author} {\bibfnamefont {X.}~\bibnamefont {{Wang}}},\ }\href
  {\doibase 10.1103/PhysRevD.99.123532} {\bibfield  {journal} {\bibinfo
  {journal} {\prd}\ }\textbf {\bibinfo {volume} {99}},\ \bibinfo {eid} {123532}
  (\bibinfo {year} {2019})},\ \Eprint {http://arxiv.org/abs/1810.11784}
  {arXiv:1810.11784 [astro-ph.CO]} \BibitemShut {NoStop}%
\bibitem [{\citenamefont {{Ruggeri}}\ \emph {et~al.}(2018)\citenamefont
  {{Ruggeri}}, \citenamefont {{Castorina}}, \citenamefont {{Carbone}},\ and\
  \citenamefont {{Sefusatti}}}]{2018JCAP...03..003R}%
  \BibitemOpen
  \bibfield  {author} {\bibinfo {author} {\bibfnamefont {R.}~\bibnamefont
  {{Ruggeri}}}, \bibinfo {author} {\bibfnamefont {E.}~\bibnamefont
  {{Castorina}}}, \bibinfo {author} {\bibfnamefont {C.}~\bibnamefont
  {{Carbone}}}, \ and\ \bibinfo {author} {\bibfnamefont {E.}~\bibnamefont
  {{Sefusatti}}},\ }\href {\doibase 10.1088/1475-7516/2018/03/003} {\bibfield
  {journal} {\bibinfo  {journal} {\jcap}\ }\textbf {\bibinfo {volume} {2018}},\
  \bibinfo {eid} {003} (\bibinfo {year} {2018})},\ \Eprint
  {http://arxiv.org/abs/1712.02334} {arXiv:1712.02334 [astro-ph.CO]}
  \BibitemShut {NoStop}%
\bibitem [{\citenamefont {{Coulton}}\ \emph {et~al.}(2019)\citenamefont
  {{Coulton}}, \citenamefont {{Liu}}, \citenamefont {{Madhavacheril}},
  \citenamefont {{B{\"o}hm}},\ and\ \citenamefont
  {{Spergel}}}]{2019JCAP...05..043C}%
  \BibitemOpen
  \bibfield  {author} {\bibinfo {author} {\bibfnamefont {W.~R.}\ \bibnamefont
  {{Coulton}}}, \bibinfo {author} {\bibfnamefont {J.}~\bibnamefont {{Liu}}},
  \bibinfo {author} {\bibfnamefont {M.~S.}\ \bibnamefont {{Madhavacheril}}},
  \bibinfo {author} {\bibfnamefont {V.}~\bibnamefont {{B{\"o}hm}}}, \ and\
  \bibinfo {author} {\bibfnamefont {D.~N.}\ \bibnamefont {{Spergel}}},\ }\href
  {\doibase 10.1088/1475-7516/2019/05/043} {\bibfield  {journal} {\bibinfo
  {journal} {\jcap}\ }\textbf {\bibinfo {volume} {2019}},\ \bibinfo {eid} {043}
  (\bibinfo {year} {2019})},\ \Eprint {http://arxiv.org/abs/1810.02374}
  {arXiv:1810.02374 [astro-ph.CO]} \BibitemShut {NoStop}%
\bibitem [{\citenamefont {{Li}}\ \emph {et~al.}(2019)\citenamefont {{Li}},
  \citenamefont {{Liu}}, \citenamefont {{Matilla}},\ and\ \citenamefont
  {{Coulton}}}]{2019PhRvD..99f3527L}%
  \BibitemOpen
  \bibfield  {author} {\bibinfo {author} {\bibfnamefont {Z.}~\bibnamefont
  {{Li}}}, \bibinfo {author} {\bibfnamefont {J.}~\bibnamefont {{Liu}}},
  \bibinfo {author} {\bibfnamefont {J.~M.~Z.}\ \bibnamefont {{Matilla}}}, \
  and\ \bibinfo {author} {\bibfnamefont {W.~R.}\ \bibnamefont {{Coulton}}},\
  }\href {\doibase 10.1103/PhysRevD.99.063527} {\bibfield  {journal} {\bibinfo
  {journal} {\prd}\ }\textbf {\bibinfo {volume} {99}},\ \bibinfo {eid} {063527}
  (\bibinfo {year} {2019})},\ \Eprint {http://arxiv.org/abs/1810.01781}
  {arXiv:1810.01781 [astro-ph.CO]} \BibitemShut {NoStop}%
\bibitem [{\citenamefont {{Marques}}\ and\ \citenamefont {{Liu et
  al.}}(2019)}]{2019JCAP...06..019M}%
  \BibitemOpen
  \bibfield  {author} {\bibinfo {author} {\bibfnamefont {G.~A.}\ \bibnamefont
  {{Marques}}}\ and\ \bibinfo {author} {\bibfnamefont {J.}~\bibnamefont {{Liu
  et al.}}},\ }\href {\doibase 10.1088/1475-7516/2019/06/019} {\bibfield
  {journal} {\bibinfo  {journal} {\jcap}\ }\textbf {\bibinfo {volume} {2019}},\
  \bibinfo {eid} {019} (\bibinfo {year} {2019})},\ \Eprint
  {http://arxiv.org/abs/1812.08206} {arXiv:1812.08206 [astro-ph.CO]}
  \BibitemShut {NoStop}%
\bibitem [{\citenamefont {Park}\ and\ \citenamefont {Kim}(2010)}]{Park_2010}%
  \BibitemOpen
  \bibfield  {author} {\bibinfo {author} {\bibfnamefont {C.}~\bibnamefont
  {Park}}\ and\ \bibinfo {author} {\bibfnamefont {Y.-R.}\ \bibnamefont {Kim}},\
  }\href {\doibase 10.1088/2041-8205/715/2/l185} {\bibfield  {journal}
  {\bibinfo  {journal} {The Astrophysical Journal}\ }\textbf {\bibinfo {volume}
  {715}},\ \bibinfo {pages} {L185} (\bibinfo {year} {2010})}\BibitemShut
  {NoStop}%
\bibitem [{\citenamefont {{Benson}}\ \emph {et~al.}(2001)\citenamefont
  {{Benson}}, \citenamefont {{Frenk}}, \citenamefont {{Baugh}}, \citenamefont
  {{Cole}},\ and\ \citenamefont {{Lacey}}}]{2001MNRAS.327.1041B}%
  \BibitemOpen
  \bibfield  {author} {\bibinfo {author} {\bibfnamefont {A.~J.}\ \bibnamefont
  {{Benson}}}, \bibinfo {author} {\bibfnamefont {C.~S.}\ \bibnamefont
  {{Frenk}}}, \bibinfo {author} {\bibfnamefont {C.~M.}\ \bibnamefont
  {{Baugh}}}, \bibinfo {author} {\bibfnamefont {S.}~\bibnamefont {{Cole}}}, \
  and\ \bibinfo {author} {\bibfnamefont {C.~G.}\ \bibnamefont {{Lacey}}},\
  }\href {\doibase 10.1046/j.1365-8711.2001.04824.x} {\bibfield  {journal}
  {\bibinfo  {journal} {\mnras}\ }\textbf {\bibinfo {volume} {327}},\ \bibinfo
  {pages} {1041} (\bibinfo {year} {2001})},\ \Eprint
  {http://arxiv.org/abs/astro-ph/0103092} {arXiv:astro-ph/0103092 [astro-ph]}
  \BibitemShut {NoStop}%
\bibitem [{\citenamefont {{Wang}}\ \emph {et~al.}(2012)\citenamefont {{Wang}},
  \citenamefont {{Chen}},\ and\ \citenamefont {{Park}}}]{2012ApJ...747...48W}%
  \BibitemOpen
  \bibfield  {author} {\bibinfo {author} {\bibfnamefont {X.}~\bibnamefont
  {{Wang}}}, \bibinfo {author} {\bibfnamefont {X.}~\bibnamefont {{Chen}}}, \
  and\ \bibinfo {author} {\bibfnamefont {C.}~\bibnamefont {{Park}}},\ }\href
  {\doibase 10.1088/0004-637X/747/1/48} {\bibfield  {journal} {\bibinfo
  {journal} {\apj}\ }\textbf {\bibinfo {volume} {747}},\ \bibinfo {eid} {48}
  (\bibinfo {year} {2012})},\ \Eprint {http://arxiv.org/abs/1010.3035}
  {arXiv:1010.3035 [astro-ph.CO]} \BibitemShut {NoStop}%
\bibitem [{\citenamefont {{Blake}}\ \emph {et~al.}(2014)\citenamefont
  {{Blake}}, \citenamefont {{James}},\ and\ \citenamefont
  {{Poole}}}]{2014MNRAS.437.2488B}%
  \BibitemOpen
  \bibfield  {author} {\bibinfo {author} {\bibfnamefont {C.}~\bibnamefont
  {{Blake}}}, \bibinfo {author} {\bibfnamefont {J.~B.}\ \bibnamefont
  {{James}}}, \ and\ \bibinfo {author} {\bibfnamefont {G.~B.}\ \bibnamefont
  {{Poole}}},\ }\href {\doibase 10.1093/mnras/stt2062} {\bibfield  {journal}
  {\bibinfo  {journal} {\mnras}\ }\textbf {\bibinfo {volume} {437}},\ \bibinfo
  {pages} {2488} (\bibinfo {year} {2014})},\ \Eprint
  {http://arxiv.org/abs/1310.6810} {arXiv:1310.6810 [astro-ph.CO]} \BibitemShut
  {NoStop}%
\bibitem [{\citenamefont {Hikage}\ \emph {et~al.}(2003)\citenamefont {Hikage},
  \citenamefont {Schmalzing},\ and\ \citenamefont
  {Buchert}}]{10.1093/pasj/55.5.911}%
  \BibitemOpen
  \bibfield  {author} {\bibinfo {author} {\bibfnamefont {C.}~\bibnamefont
  {Hikage}}, \bibinfo {author} {\bibfnamefont {J.}~\bibnamefont {Schmalzing}},
  \ and\ \bibinfo {author} {\bibfnamefont {T.~e.~a.}\ \bibnamefont {Buchert}},\
  }\href {\doibase 10.1093/pasj/55.5.911} {\bibfield  {journal} {\bibinfo
  {journal} {Publications of the Astronomical Society of Japan}\ }\textbf
  {\bibinfo {volume} {55}},\ \bibinfo {pages} {911} (\bibinfo {year} {2003})},\
  \Eprint
  {http://arxiv.org/abs/http://oup.prod.sis.lan/pasj/article-pdf/55/5/911/17447854/pasj55-0911.pdf}
  {http://oup.prod.sis.lan/pasj/article-pdf/55/5/911/17447854/pasj55-0911.pdf}
  \BibitemShut {NoStop}%
\bibitem [{\citenamefont {{Massara}}\ \emph {et~al.}(2015)\citenamefont
  {{Massara}}, \citenamefont {{Villaescusa-Navarro}}, \citenamefont {{Viel}},\
  and\ \citenamefont {{Sutter}}}]{2015JCAP...11..018M}%
  \BibitemOpen
  \bibfield  {author} {\bibinfo {author} {\bibfnamefont {E.}~\bibnamefont
  {{Massara}}}, \bibinfo {author} {\bibfnamefont {F.}~\bibnamefont
  {{Villaescusa-Navarro}}}, \bibinfo {author} {\bibfnamefont {M.}~\bibnamefont
  {{Viel}}}, \ and\ \bibinfo {author} {\bibfnamefont {P.~M.}\ \bibnamefont
  {{Sutter}}},\ }\href {\doibase 10.1088/1475-7516/2015/11/018} {\bibfield
  {journal} {\bibinfo  {journal} {\jcap}\ }\textbf {\bibinfo {volume} {2015}},\
  \bibinfo {eid} {018} (\bibinfo {year} {2015})},\ \Eprint
  {http://arxiv.org/abs/1506.03088} {arXiv:1506.03088 [astro-ph.CO]}
  \BibitemShut {NoStop}%
\bibitem [{\citenamefont {{Kreisch}}\ \emph {et~al.}(2019)\citenamefont
  {{Kreisch}}, \citenamefont {{Pisani}}, \citenamefont {{Carbone}},
  \citenamefont {{Liu}}, \citenamefont {{Hawken}}, \citenamefont {{Massara}},
  \citenamefont {{Spergel}},\ and\ \citenamefont
  {{Wandelt}}}]{2019MNRAS.488.4413K}%
  \BibitemOpen
  \bibfield  {author} {\bibinfo {author} {\bibfnamefont {C.~D.}\ \bibnamefont
  {{Kreisch}}}, \bibinfo {author} {\bibfnamefont {A.}~\bibnamefont {{Pisani}}},
  \bibinfo {author} {\bibfnamefont {C.}~\bibnamefont {{Carbone}}}, \bibinfo
  {author} {\bibfnamefont {J.}~\bibnamefont {{Liu}}}, \bibinfo {author}
  {\bibfnamefont {A.~J.}\ \bibnamefont {{Hawken}}}, \bibinfo {author}
  {\bibfnamefont {E.}~\bibnamefont {{Massara}}}, \bibinfo {author}
  {\bibfnamefont {D.~N.}\ \bibnamefont {{Spergel}}}, \ and\ \bibinfo {author}
  {\bibfnamefont {B.~D.}\ \bibnamefont {{Wandelt}}},\ }\href {\doibase
  10.1093/mnras/stz1944} {\bibfield  {journal} {\bibinfo  {journal} {\mnras}\
  }\textbf {\bibinfo {volume} {488}},\ \bibinfo {pages} {4413} (\bibinfo {year}
  {2019})},\ \Eprint {http://arxiv.org/abs/1808.07464} {arXiv:1808.07464
  [astro-ph.CO]} \BibitemShut {NoStop}%
\bibitem [{\citenamefont {{Ichiki}}\ and\ \citenamefont
  {{Takada}}(2012)}]{2012PhRvD..85f3521I}%
  \BibitemOpen
  \bibfield  {author} {\bibinfo {author} {\bibfnamefont {K.}~\bibnamefont
  {{Ichiki}}}\ and\ \bibinfo {author} {\bibfnamefont {M.}~\bibnamefont
  {{Takada}}},\ }\href {\doibase 10.1103/PhysRevD.85.063521} {\bibfield
  {journal} {\bibinfo  {journal} {\prd}\ }\textbf {\bibinfo {volume} {85}},\
  \bibinfo {eid} {063521} (\bibinfo {year} {2012})},\ \Eprint
  {http://arxiv.org/abs/1108.4688} {arXiv:1108.4688 [astro-ph.CO]} \BibitemShut
  {NoStop}%
\bibitem [{\citenamefont {{Costanzi}}\ and\ \citenamefont {{Villaescusa-Navarro
  et al.}}(2013)}]{2013JCAP...12..012C}%
  \BibitemOpen
  \bibfield  {author} {\bibinfo {author} {\bibfnamefont {M.}~\bibnamefont
  {{Costanzi}}}\ and\ \bibinfo {author} {\bibfnamefont {F.}~\bibnamefont
  {{Villaescusa-Navarro et al.}}},\ }\href {\doibase
  10.1088/1475-7516/2013/12/012} {\bibfield  {journal} {\bibinfo  {journal}
  {\jcap}\ }\textbf {\bibinfo {volume} {2013}},\ \bibinfo {eid} {012} (\bibinfo
  {year} {2013})},\ \Eprint {http://arxiv.org/abs/1311.1514} {arXiv:1311.1514
  [astro-ph.CO]} \BibitemShut {NoStop}%
\bibitem [{\citenamefont {{Kratochvil}}\ \emph {et~al.}(2012)\citenamefont
  {{Kratochvil}}, \citenamefont {{Lim}},\ and\ \citenamefont {{Wang et
  al.}}}]{2012PhRvD..85j3513K}%
  \BibitemOpen
  \bibfield  {author} {\bibinfo {author} {\bibfnamefont {J.~M.}\ \bibnamefont
  {{Kratochvil}}}, \bibinfo {author} {\bibfnamefont {E.~A.}\ \bibnamefont
  {{Lim}}}, \ and\ \bibinfo {author} {\bibfnamefont {S.}~\bibnamefont {{Wang et
  al.}}},\ }\href {\doibase 10.1103/PhysRevD.85.103513} {\bibfield  {journal}
  {\bibinfo  {journal} {\prd}\ }\textbf {\bibinfo {volume} {85}},\ \bibinfo
  {eid} {103513} (\bibinfo {year} {2012})},\ \Eprint
  {http://arxiv.org/abs/1109.6334} {arXiv:1109.6334 [astro-ph.CO]} \BibitemShut
  {NoStop}%
\bibitem [{\citenamefont {{Shirasaki}}\ \emph {et~al.}(2017)\citenamefont
  {{Shirasaki}}, \citenamefont {{Nishimichi}}, \citenamefont {{Li}},\ and\
  \citenamefont {{Higuchi}}}]{2017MNRAS.466.2402S}%
  \BibitemOpen
  \bibfield  {author} {\bibinfo {author} {\bibfnamefont {M.}~\bibnamefont
  {{Shirasaki}}}, \bibinfo {author} {\bibfnamefont {T.}~\bibnamefont
  {{Nishimichi}}}, \bibinfo {author} {\bibfnamefont {B.}~\bibnamefont {{Li}}},
  \ and\ \bibinfo {author} {\bibfnamefont {Y.}~\bibnamefont {{Higuchi}}},\
  }\href {\doibase 10.1093/mnras/stw3254} {\bibfield  {journal} {\bibinfo
  {journal} {\mnras}\ }\textbf {\bibinfo {volume} {466}},\ \bibinfo {pages}
  {2402} (\bibinfo {year} {2017})},\ \Eprint {http://arxiv.org/abs/1610.03600}
  {arXiv:1610.03600 [astro-ph.CO]} \BibitemShut {NoStop}%
\bibitem [{\citenamefont {{Peel}}\ \emph {et~al.}(2018)\citenamefont {{Peel}},
  \citenamefont {{Pettorino}}, \citenamefont {{Giocoli}}, \citenamefont
  {{Starck}},\ and\ \citenamefont {{Baldi}}}]{2018A&A...619A..38P}%
  \BibitemOpen
  \bibfield  {author} {\bibinfo {author} {\bibfnamefont {A.}~\bibnamefont
  {{Peel}}}, \bibinfo {author} {\bibfnamefont {V.}~\bibnamefont {{Pettorino}}},
  \bibinfo {author} {\bibfnamefont {C.}~\bibnamefont {{Giocoli}}}, \bibinfo
  {author} {\bibfnamefont {J.-L.}\ \bibnamefont {{Starck}}}, \ and\ \bibinfo
  {author} {\bibfnamefont {M.}~\bibnamefont {{Baldi}}},\ }\href {\doibase
  10.1051/0004-6361/201833481} {\bibfield  {journal} {\bibinfo  {journal}
  {\aap}\ }\textbf {\bibinfo {volume} {619}},\ \bibinfo {eid} {A38} (\bibinfo
  {year} {2018})},\ \Eprint {http://arxiv.org/abs/1805.05146} {arXiv:1805.05146
  [astro-ph.CO]} \BibitemShut {NoStop}%
\bibitem [{\citenamefont {{Vafaei}}\ \emph {et~al.}(2010)\citenamefont
  {{Vafaei}}, \citenamefont {{Lu}}, \citenamefont {{van Waerbeke}},
  \citenamefont {{Semboloni}}, \citenamefont {{Heymans}},\ and\ \citenamefont
  {{Pen}}}]{2010APh....32..340V}%
  \BibitemOpen
  \bibfield  {author} {\bibinfo {author} {\bibfnamefont {S.}~\bibnamefont
  {{Vafaei}}}, \bibinfo {author} {\bibfnamefont {T.}~\bibnamefont {{Lu}}},
  \bibinfo {author} {\bibfnamefont {L.}~\bibnamefont {{van Waerbeke}}},
  \bibinfo {author} {\bibfnamefont {E.}~\bibnamefont {{Semboloni}}}, \bibinfo
  {author} {\bibfnamefont {C.}~\bibnamefont {{Heymans}}}, \ and\ \bibinfo
  {author} {\bibfnamefont {U.-L.}\ \bibnamefont {{Pen}}},\ }\href {\doibase
  10.1016/j.astropartphys.2009.10.003} {\bibfield  {journal} {\bibinfo
  {journal} {Astroparticle Physics}\ }\textbf {\bibinfo {volume} {32}},\
  \bibinfo {pages} {340} (\bibinfo {year} {2010})},\ \Eprint
  {http://arxiv.org/abs/0905.3726} {arXiv:0905.3726 [astro-ph.CO]} \BibitemShut
  {NoStop}%
\bibitem [{\citenamefont {{Hahn}}\ \emph {et~al.}(2019)\citenamefont {{Hahn}},
  \citenamefont {{Francisco}}, \citenamefont {{Emanuele}},\ and\ \citenamefont
  {{Roman}}}]{2019arXiv190911107H}%
  \BibitemOpen
  \bibfield  {author} {\bibinfo {author} {\bibfnamefont {C.}~\bibnamefont
  {{Hahn}}}, \bibinfo {author} {\bibfnamefont {V.-N.}\ \bibnamefont
  {{Francisco}}}, \bibinfo {author} {\bibfnamefont {C.}~\bibnamefont
  {{Emanuele}}}, \ and\ \bibinfo {author} {\bibfnamefont {S.}~\bibnamefont
  {{Roman}}},\ }\href@noop {} {\bibfield  {journal} {\bibinfo  {journal} {arXiv
  e-prints}\ ,\ \bibinfo {eid} {arXiv:1909.11107}} (\bibinfo {year} {2019})},\
  \Eprint {http://arxiv.org/abs/1909.11107} {arXiv:1909.11107 [astro-ph.CO]}
  \BibitemShut {NoStop}%
\bibitem [{\citenamefont {{Xu}}\ \emph {et~al.}(2019)\citenamefont {{Xu}},
  \citenamefont {{Cisewski-Kehe}}, \citenamefont {{Green}},\ and\ \citenamefont
  {{Nagai}}}]{2019A&C....27...34X}%
  \BibitemOpen
  \bibfield  {author} {\bibinfo {author} {\bibfnamefont {X.}~\bibnamefont
  {{Xu}}}, \bibinfo {author} {\bibfnamefont {J.}~\bibnamefont
  {{Cisewski-Kehe}}}, \bibinfo {author} {\bibfnamefont {S.~B.}\ \bibnamefont
  {{Green}}}, \ and\ \bibinfo {author} {\bibfnamefont {D.}~\bibnamefont
  {{Nagai}}},\ }\href {\doibase 10.1016/j.ascom.2019.02.003} {\bibfield
  {journal} {\bibinfo  {journal} {Astronomy and Computing}\ }\textbf {\bibinfo
  {volume} {27}},\ \bibinfo {eid} {34} (\bibinfo {year} {2019})},\ \Eprint
  {http://arxiv.org/abs/1811.08450} {arXiv:1811.08450 [astro-ph.CO]}
  \BibitemShut {NoStop}%
\bibitem [{\citenamefont {{Mecke}}\ \emph {et~al.}(1994)\citenamefont
  {{Mecke}}, \citenamefont {{Buchert}},\ and\ \citenamefont
  {{Wagner}}}]{1994A&A...288..697M}%
  \BibitemOpen
  \bibfield  {author} {\bibinfo {author} {\bibfnamefont {K.~R.}\ \bibnamefont
  {{Mecke}}}, \bibinfo {author} {\bibfnamefont {T.}~\bibnamefont {{Buchert}}},
  \ and\ \bibinfo {author} {\bibfnamefont {H.}~\bibnamefont {{Wagner}}},\
  }\href@noop {} {\bibfield  {journal} {\bibinfo  {journal} {\aap}\ }\textbf
  {\bibinfo {volume} {288}},\ \bibinfo {pages} {697} (\bibinfo {year}
  {1994})},\ \Eprint {http://arxiv.org/abs/astro-ph/9312028}
  {arXiv:astro-ph/9312028 [astro-ph]} \BibitemShut {NoStop}%
\bibitem [{\citenamefont {{Ducout}}\ \emph {et~al.}(2013)\citenamefont
  {{Ducout}}, \citenamefont {{Bouchet}}, \citenamefont {{Colombi}},
  \citenamefont {{Pogosyan}},\ and\ \citenamefont
  {{Prunet}}}]{2013MNRAS.429.2104D}%
  \BibitemOpen
  \bibfield  {author} {\bibinfo {author} {\bibfnamefont {A.}~\bibnamefont
  {{Ducout}}}, \bibinfo {author} {\bibfnamefont {F.~R.}\ \bibnamefont
  {{Bouchet}}}, \bibinfo {author} {\bibfnamefont {S.}~\bibnamefont
  {{Colombi}}}, \bibinfo {author} {\bibfnamefont {D.}~\bibnamefont
  {{Pogosyan}}}, \ and\ \bibinfo {author} {\bibfnamefont {S.}~\bibnamefont
  {{Prunet}}},\ }\href {\doibase 10.1093/mnras/sts483} {\bibfield  {journal}
  {\bibinfo  {journal} {\mnras}\ }\textbf {\bibinfo {volume} {429}},\ \bibinfo
  {pages} {2104} (\bibinfo {year} {2013})},\ \Eprint
  {http://arxiv.org/abs/1209.1223} {arXiv:1209.1223 [astro-ph.CO]} \BibitemShut
  {NoStop}%
\bibitem [{\citenamefont {{Planck Collaboration}}\ and\ \citenamefont {{Ade et
  al.}}(2016)}]{2016A&A...594A..17P}%
  \BibitemOpen
  \bibfield  {author} {\bibinfo {author} {\bibnamefont {{Planck
  Collaboration}}}\ and\ \bibinfo {author} {\bibfnamefont {P.~A.~R.}\
  \bibnamefont {{Ade et al.}}},\ }\href {\doibase 10.1051/0004-6361/201525836}
  {\bibfield  {journal} {\bibinfo  {journal} {\aap}\ }\textbf {\bibinfo
  {volume} {594}},\ \bibinfo {eid} {A17} (\bibinfo {year} {2016})},\ \Eprint
  {http://arxiv.org/abs/1502.01592} {arXiv:1502.01592 [astro-ph.CO]}
  \BibitemShut {NoStop}%
\bibitem [{\citenamefont {Hadwiger}(1957)}]{Hadwiger1957Vorlesungen}%
  \BibitemOpen
  \bibfield  {author} {\bibinfo {author} {\bibfnamefont {H.}~\bibnamefont
  {Hadwiger}},\ }\href@noop {} {\emph {\bibinfo {title} {Vorlesungen \"{u}ber
  Inhalt, Oberfl\"{a}che und Isoperimetrie}}}\ (\bibinfo {year} {Springer,
  Berlin, 1957})\BibitemShut {NoStop}%
\bibitem [{\citenamefont {{Gott}}\ \emph {et~al.}(1986)\citenamefont {{Gott}},
  \citenamefont {{Melott}},\ and\ \citenamefont
  {{Dickinson}}}]{1986ApJ...306..341G}%
  \BibitemOpen
  \bibfield  {author} {\bibinfo {author} {\bibfnamefont {I.}~\bibnamefont
  {{Gott}}, \bibfnamefont {J.~Richard}}, \bibinfo {author} {\bibfnamefont
  {A.~L.}\ \bibnamefont {{Melott}}}, \ and\ \bibinfo {author} {\bibfnamefont
  {M.}~\bibnamefont {{Dickinson}}},\ }\href {\doibase 10.1086/164347}
  {\bibfield  {journal} {\bibinfo  {journal} {\apj}\ }\textbf {\bibinfo
  {volume} {306}},\ \bibinfo {pages} {341} (\bibinfo {year}
  {1986})}\BibitemShut {NoStop}%
\bibitem [{\citenamefont {Schmalzing}\ and\ \citenamefont
  {Buchert}(1997)}]{Schmalzing_1997}%
  \BibitemOpen
  \bibfield  {author} {\bibinfo {author} {\bibfnamefont {J.}~\bibnamefont
  {Schmalzing}}\ and\ \bibinfo {author} {\bibfnamefont {T.}~\bibnamefont
  {Buchert}},\ }\href {\doibase 10.1086/310680} {\bibfield  {journal} {\bibinfo
   {journal} {The Astrophysical Journal}\ }\textbf {\bibinfo {volume} {482}},\
  \bibinfo {pages} {L1} (\bibinfo {year} {1997})}\BibitemShut {NoStop}%
\bibitem [{\citenamefont {Melott}(1990)}]{MELOTT19901}%
  \BibitemOpen
  \bibfield  {author} {\bibinfo {author} {\bibfnamefont {A.~L.}\ \bibnamefont
  {Melott}},\ }\href {\doibase https://doi.org/10.1016/0370-1573(90)90162-U}
  {\bibfield  {journal} {\bibinfo  {journal} {Physics Reports}\ }\textbf
  {\bibinfo {volume} {193}},\ \bibinfo {pages} {1 } (\bibinfo {year}
  {1990})}\BibitemShut {NoStop}%
\bibitem [{\citenamefont {{F{\"u}hrer}}\ and\ \citenamefont
  {{Wong}}(2015)}]{2015JCAP...03..046F}%
  \BibitemOpen
  \bibfield  {author} {\bibinfo {author} {\bibfnamefont {F.}~\bibnamefont
  {{F{\"u}hrer}}}\ and\ \bibinfo {author} {\bibfnamefont {Y.~Y.~Y.}\
  \bibnamefont {{Wong}}},\ }\href {\doibase 10.1088/1475-7516/2015/03/046}
  {\bibfield  {journal} {\bibinfo  {journal} {\jcap}\ }\textbf {\bibinfo
  {volume} {2015}},\ \bibinfo {eid} {046} (\bibinfo {year} {2015})},\ \Eprint
  {http://arxiv.org/abs/1412.2764} {arXiv:1412.2764 [astro-ph.CO]} \BibitemShut
  {NoStop}%
\bibitem [{\citenamefont {{Brandbyge}}\ and\ \citenamefont
  {{Hannestad}}(2009)}]{2009JCAP...05..002B}%
  \BibitemOpen
  \bibfield  {author} {\bibinfo {author} {\bibfnamefont {J.}~\bibnamefont
  {{Brandbyge}}}\ and\ \bibinfo {author} {\bibfnamefont {S.}~\bibnamefont
  {{Hannestad}}},\ }\href {\doibase 10.1088/1475-7516/2009/05/002} {\bibfield
  {journal} {\bibinfo  {journal} {\jcap}\ }\textbf {\bibinfo {volume} {2009}},\
  \bibinfo {eid} {002} (\bibinfo {year} {2009})},\ \Eprint
  {http://arxiv.org/abs/0812.3149} {arXiv:0812.3149 [astro-ph]} \BibitemShut
  {NoStop}%
\bibitem [{\citenamefont {{Ali-Ha{\"\i}moud}}\ and\ \citenamefont
  {{Bird}}(2013)}]{2013MNRAS.428.3375A}%
  \BibitemOpen
  \bibfield  {author} {\bibinfo {author} {\bibfnamefont {Y.}~\bibnamefont
  {{Ali-Ha{\"\i}moud}}}\ and\ \bibinfo {author} {\bibfnamefont
  {S.}~\bibnamefont {{Bird}}},\ }\href {\doibase 10.1093/mnras/sts286}
  {\bibfield  {journal} {\bibinfo  {journal} {\mnras}\ }\textbf {\bibinfo
  {volume} {428}},\ \bibinfo {pages} {3375} (\bibinfo {year} {2013})},\ \Eprint
  {http://arxiv.org/abs/1209.0461} {arXiv:1209.0461 [astro-ph.CO]} \BibitemShut
  {NoStop}%
\bibitem [{\citenamefont {{Brandbyge}}\ and\ \citenamefont
  {{Hannestad}}(2010)}]{2010JCAP...01..021B}%
  \BibitemOpen
  \bibfield  {author} {\bibinfo {author} {\bibfnamefont {J.}~\bibnamefont
  {{Brandbyge}}}\ and\ \bibinfo {author} {\bibfnamefont {S.}~\bibnamefont
  {{Hannestad}}},\ }\href {\doibase 10.1088/1475-7516/2010/01/021} {\bibfield
  {journal} {\bibinfo  {journal} {\jcap}\ }\textbf {\bibinfo {volume} {2010}},\
  \bibinfo {eid} {021} (\bibinfo {year} {2010})},\ \Eprint
  {http://arxiv.org/abs/0908.1969} {arXiv:0908.1969 [astro-ph.CO]} \BibitemShut
  {NoStop}%
\bibitem [{\citenamefont {{Bird}}\ \emph {et~al.}(2018)\citenamefont {{Bird}},
  \citenamefont {{Ali-Ha{\"\i}moud}}, \citenamefont {{Feng}},\ and\
  \citenamefont {{Liu}}}]{2018MNRAS.481.1486B}%
  \BibitemOpen
  \bibfield  {author} {\bibinfo {author} {\bibfnamefont {S.}~\bibnamefont
  {{Bird}}}, \bibinfo {author} {\bibfnamefont {Y.}~\bibnamefont
  {{Ali-Ha{\"\i}moud}}}, \bibinfo {author} {\bibfnamefont {Y.}~\bibnamefont
  {{Feng}}}, \ and\ \bibinfo {author} {\bibfnamefont {J.}~\bibnamefont
  {{Liu}}},\ }\href {\doibase 10.1093/mnras/sty2376} {\bibfield  {journal}
  {\bibinfo  {journal} {\mnras}\ }\textbf {\bibinfo {volume} {481}},\ \bibinfo
  {pages} {1486} (\bibinfo {year} {2018})},\ \Eprint
  {http://arxiv.org/abs/1803.09854} {arXiv:1803.09854 [astro-ph.CO]}
  \BibitemShut {NoStop}%
\bibitem [{\citenamefont {{Banerjee}}\ and\ \citenamefont
  {{Dalal}}(2016)}]{2016JCAP...11..015B}%
  \BibitemOpen
  \bibfield  {author} {\bibinfo {author} {\bibfnamefont {A.}~\bibnamefont
  {{Banerjee}}}\ and\ \bibinfo {author} {\bibfnamefont {N.}~\bibnamefont
  {{Dalal}}},\ }\href {\doibase 10.1088/1475-7516/2016/11/015} {\bibfield
  {journal} {\bibinfo  {journal} {\jcap}\ }\textbf {\bibinfo {volume} {2016}},\
  \bibinfo {eid} {015} (\bibinfo {year} {2016})},\ \Eprint
  {http://arxiv.org/abs/1606.06167} {arXiv:1606.06167 [astro-ph.CO]}
  \BibitemShut {NoStop}%
\bibitem [{\citenamefont {{Inman}}\ and\ \citenamefont
  {{Pen}}(2017)}]{2017PhRvD..95f3535I}%
  \BibitemOpen
  \bibfield  {author} {\bibinfo {author} {\bibfnamefont {D.}~\bibnamefont
  {{Inman}}}\ and\ \bibinfo {author} {\bibfnamefont {U.-L.}\ \bibnamefont
  {{Pen}}},\ }\href {\doibase 10.1103/PhysRevD.95.063535} {\bibfield  {journal}
  {\bibinfo  {journal} {\prd}\ }\textbf {\bibinfo {volume} {95}},\ \bibinfo
  {eid} {063535} (\bibinfo {year} {2017})},\ \Eprint
  {http://arxiv.org/abs/1609.09469} {arXiv:1609.09469 [astro-ph.CO]}
  \BibitemShut {NoStop}%
\bibitem [{\citenamefont {{Emberson}}\ \emph {et~al.}(2017)\citenamefont
  {{Emberson}}, \citenamefont {{Yu}},\ and\ \citenamefont {{Inman et
  al.}}}]{2017RAA....17...85E}%
  \BibitemOpen
  \bibfield  {author} {\bibinfo {author} {\bibfnamefont {J.~D.}\ \bibnamefont
  {{Emberson}}}, \bibinfo {author} {\bibfnamefont {H.-R.}\ \bibnamefont
  {{Yu}}}, \ and\ \bibinfo {author} {\bibfnamefont {D.}~\bibnamefont {{Inman et
  al.}}},\ }\href {\doibase 10.1088/1674-4527/17/8/85} {\bibfield  {journal}
  {\bibinfo  {journal} {Research in Astronomy and Astrophysics}\ }\textbf
  {\bibinfo {volume} {17}},\ \bibinfo {eid} {085} (\bibinfo {year} {2017})},\
  \Eprint {http://arxiv.org/abs/1611.01545} {arXiv:1611.01545 [astro-ph.CO]}
  \BibitemShut {NoStop}%
\bibitem [{\citenamefont {{Harnois-D{\'e}raps}}\ \emph
  {et~al.}(2013)\citenamefont {{Harnois-D{\'e}raps}}, \citenamefont {{Pen}},\
  and\ \citenamefont {{Iliev et al.}}}]{2013MNRAS.436..540H}%
  \BibitemOpen
  \bibfield  {author} {\bibinfo {author} {\bibfnamefont {J.}~\bibnamefont
  {{Harnois-D{\'e}raps}}}, \bibinfo {author} {\bibfnamefont {U.-L.}\
  \bibnamefont {{Pen}}}, \ and\ \bibinfo {author} {\bibfnamefont {I.~T.}\
  \bibnamefont {{Iliev et al.}}},\ }\href {\doibase 10.1093/mnras/stt1591}
  {\bibfield  {journal} {\bibinfo  {journal} {\mnras}\ }\textbf {\bibinfo
  {volume} {436}},\ \bibinfo {pages} {540} (\bibinfo {year} {2013})},\ \Eprint
  {http://arxiv.org/abs/1208.5098} {arXiv:1208.5098 [astro-ph.CO]} \BibitemShut
  {NoStop}%
\bibitem [{\citenamefont {{Inman}}\ \emph {et~al.}(2017)\citenamefont
  {{Inman}}, \citenamefont {{Yu}},\ and\ \citenamefont {{Zhu et
  al.}}}]{2017PhRvD..95h3518I}%
  \BibitemOpen
  \bibfield  {author} {\bibinfo {author} {\bibfnamefont {D.}~\bibnamefont
  {{Inman}}}, \bibinfo {author} {\bibfnamefont {H.-R.}\ \bibnamefont {{Yu}}}, \
  and\ \bibinfo {author} {\bibfnamefont {H.-M.}\ \bibnamefont {{Zhu et al.}}},\
  }\href {\doibase 10.1103/PhysRevD.95.083518} {\bibfield  {journal} {\bibinfo
  {journal} {\prd}\ }\textbf {\bibinfo {volume} {95}},\ \bibinfo {eid} {083518}
  (\bibinfo {year} {2017})}\BibitemShut {NoStop}%
\bibitem [{\citenamefont {{Yu}}\ \emph {et~al.}(2017)\citenamefont {{Yu}},
  \citenamefont {{Emberson}},\ and\ \citenamefont {{Inman et
  al.}}}]{2017NatAs...1E.143Y}%
  \BibitemOpen
  \bibfield  {author} {\bibinfo {author} {\bibfnamefont {H.-R.}\ \bibnamefont
  {{Yu}}}, \bibinfo {author} {\bibfnamefont {J.~D.}\ \bibnamefont
  {{Emberson}}}, \ and\ \bibinfo {author} {\bibfnamefont {D.}~\bibnamefont
  {{Inman et al.}}},\ }\href {\doibase 10.1038/s41550-017-0143} {\bibfield
  {journal} {\bibinfo  {journal} {Nature Astronomy}\ }\textbf {\bibinfo
  {volume} {1}},\ \bibinfo {eid} {0143} (\bibinfo {year} {2017})},\ \Eprint
  {http://arxiv.org/abs/1609.08968} {arXiv:1609.08968 [astro-ph.CO]}
  \BibitemShut {NoStop}%
\bibitem [{\citenamefont {{Blas}}\ \emph {et~al.}(2011)\citenamefont {{Blas}},
  \citenamefont {{Lesgourgues}},\ and\ \citenamefont
  {{Tram}}}]{2011JCAP...07..034B}%
  \BibitemOpen
  \bibfield  {author} {\bibinfo {author} {\bibfnamefont {D.}~\bibnamefont
  {{Blas}}}, \bibinfo {author} {\bibfnamefont {J.}~\bibnamefont
  {{Lesgourgues}}}, \ and\ \bibinfo {author} {\bibfnamefont {T.}~\bibnamefont
  {{Tram}}},\ }\href {\doibase 10.1088/1475-7516/2011/07/034} {\bibfield
  {journal} {\bibinfo  {journal} {\jcap}\ }\textbf {\bibinfo {volume} {2011}},\
  \bibinfo {eid} {034} (\bibinfo {year} {2011})},\ \Eprint
  {http://arxiv.org/abs/1104.2933} {arXiv:1104.2933 [astro-ph.CO]} \BibitemShut
  {NoStop}%
\bibitem [{\citenamefont {{Coles}}\ and\ \citenamefont
  {{Jones}}(1991)}]{1991MNRAS.248....1C}%
  \BibitemOpen
  \bibfield  {author} {\bibinfo {author} {\bibfnamefont {P.}~\bibnamefont
  {{Coles}}}\ and\ \bibinfo {author} {\bibfnamefont {B.}~\bibnamefont
  {{Jones}}},\ }\href {\doibase 10.1093/mnras/248.1.1} {\bibfield  {journal}
  {\bibinfo  {journal} {\mnras}\ }\textbf {\bibinfo {volume} {248}},\ \bibinfo
  {pages} {1} (\bibinfo {year} {1991})}\BibitemShut {NoStop}%
\bibitem [{\citenamefont {Press}\ \emph {et~al.}(1992)\citenamefont {Press},
  \citenamefont {Flannery}, \citenamefont {Teukolsky},\ and\ \citenamefont
  {Vetterling}}]{numerical.book}%
  \BibitemOpen
  \bibfield  {author} {\bibinfo {author} {\bibfnamefont {W.~H.}\ \bibnamefont
  {Press}}, \bibinfo {author} {\bibfnamefont {B.~P.}\ \bibnamefont {Flannery}},
  \bibinfo {author} {\bibfnamefont {S.~A.}\ \bibnamefont {Teukolsky}}, \ and\
  \bibinfo {author} {\bibfnamefont {W.~T.}\ \bibnamefont {Vetterling}},\
  }\href@noop {} {\emph {\bibinfo {title} {{Numerical Recipes in FORTRAN: The
  Art of Scientific Computing, 2nd ed. p. 465}}}}\ (\bibinfo {year}
  {1992})\BibitemShut {NoStop}%
\bibitem [{\citenamefont {{Massey}}\ \emph {et~al.}(2007)\citenamefont
  {{Massey}}, \citenamefont {{Rhodes}}, \citenamefont {{Ellis}}, \citenamefont
  {{Scoville}}, \citenamefont {{Leauthaud}}, \citenamefont {{Finoguenov}},
  \citenamefont {{Capak}}, \citenamefont {{Bacon}}, \citenamefont {{Aussel}},
  \citenamefont {{Kneib}}, \citenamefont {{Koekemoer}}, \citenamefont
  {{McCracken}}, \citenamefont {{Mobasher}}, \citenamefont {{Pires}},
  \citenamefont {{Refregier}}, \citenamefont {{Sasaki}}, \citenamefont
  {{Starck}}, \citenamefont {{Taniguchi}}, \citenamefont {{Taylor}},\ and\
  \citenamefont {{Taylor}}}]{2007Natur.445..286M}%
  \BibitemOpen
  \bibfield  {author} {\bibinfo {author} {\bibfnamefont {R.}~\bibnamefont
  {{Massey}}}, \bibinfo {author} {\bibfnamefont {J.}~\bibnamefont {{Rhodes}}},
  \bibinfo {author} {\bibfnamefont {R.}~\bibnamefont {{Ellis}}}, \bibinfo
  {author} {\bibfnamefont {N.}~\bibnamefont {{Scoville}}}, \bibinfo {author}
  {\bibfnamefont {A.}~\bibnamefont {{Leauthaud}}}, \bibinfo {author}
  {\bibfnamefont {A.}~\bibnamefont {{Finoguenov}}}, \bibinfo {author}
  {\bibfnamefont {P.}~\bibnamefont {{Capak}}}, \bibinfo {author} {\bibfnamefont
  {D.}~\bibnamefont {{Bacon}}}, \bibinfo {author} {\bibfnamefont
  {H.}~\bibnamefont {{Aussel}}}, \bibinfo {author} {\bibfnamefont {J.-P.}\
  \bibnamefont {{Kneib}}}, \bibinfo {author} {\bibfnamefont {A.}~\bibnamefont
  {{Koekemoer}}}, \bibinfo {author} {\bibfnamefont {H.}~\bibnamefont
  {{McCracken}}}, \bibinfo {author} {\bibfnamefont {B.}~\bibnamefont
  {{Mobasher}}}, \bibinfo {author} {\bibfnamefont {S.}~\bibnamefont {{Pires}}},
  \bibinfo {author} {\bibfnamefont {A.}~\bibnamefont {{Refregier}}}, \bibinfo
  {author} {\bibfnamefont {S.}~\bibnamefont {{Sasaki}}}, \bibinfo {author}
  {\bibfnamefont {J.-L.}\ \bibnamefont {{Starck}}}, \bibinfo {author}
  {\bibfnamefont {Y.}~\bibnamefont {{Taniguchi}}}, \bibinfo {author}
  {\bibfnamefont {A.}~\bibnamefont {{Taylor}}}, \ and\ \bibinfo {author}
  {\bibfnamefont {J.}~\bibnamefont {{Taylor}}},\ }\href {\doibase
  10.1038/nature05497} {\bibfield  {journal} {\bibinfo  {journal} {\nat}\
  }\textbf {\bibinfo {volume} {445}},\ \bibinfo {pages} {286} (\bibinfo {year}
  {2007})},\ \Eprint {http://arxiv.org/abs/astro-ph/0701594}
  {arXiv:astro-ph/0701594 [astro-ph]} \BibitemShut {NoStop}%
\bibitem [{\citenamefont {{Sheth}}\ and\ \citenamefont {{van de
  Weygaert}}(2004)}]{2004MNRAS.350..517S}%
  \BibitemOpen
  \bibfield  {author} {\bibinfo {author} {\bibfnamefont {R.~K.}\ \bibnamefont
  {{Sheth}}}\ and\ \bibinfo {author} {\bibfnamefont {R.}~\bibnamefont {{van de
  Weygaert}}},\ }\href {\doibase 10.1111/j.1365-2966.2004.07661.x} {\bibfield
  {journal} {\bibinfo  {journal} {\mnras}\ }\textbf {\bibinfo {volume} {350}},\
  \bibinfo {pages} {517} (\bibinfo {year} {2004})},\ \Eprint
  {http://arxiv.org/abs/astro-ph/0311260} {arXiv:astro-ph/0311260 [astro-ph]}
  \BibitemShut {NoStop}%
\bibitem [{\citenamefont {{Agarwal}}\ and\ \citenamefont
  {{Feldman}}(2011)}]{2011MNRAS.410.1647A}%
  \BibitemOpen
  \bibfield  {author} {\bibinfo {author} {\bibfnamefont {S.}~\bibnamefont
  {{Agarwal}}}\ and\ \bibinfo {author} {\bibfnamefont {H.~A.}\ \bibnamefont
  {{Feldman}}},\ }\href {\doibase 10.1111/j.1365-2966.2010.17546.x} {\bibfield
  {journal} {\bibinfo  {journal} {\mnras}\ }\textbf {\bibinfo {volume} {410}},\
  \bibinfo {pages} {1647} (\bibinfo {year} {2011})},\ \Eprint
  {http://arxiv.org/abs/1006.0689} {arXiv:1006.0689 [astro-ph.CO]} \BibitemShut
  {NoStop}%
\bibitem [{\citenamefont {{Massara}}\ \emph {et~al.}(2014)\citenamefont
  {{Massara}}, \citenamefont {{Villaescusa-Navarro}},\ and\ \citenamefont
  {{Viel}}}]{2014JCAP...12..053M}%
  \BibitemOpen
  \bibfield  {author} {\bibinfo {author} {\bibfnamefont {E.}~\bibnamefont
  {{Massara}}}, \bibinfo {author} {\bibfnamefont {F.}~\bibnamefont
  {{Villaescusa-Navarro}}}, \ and\ \bibinfo {author} {\bibfnamefont
  {M.}~\bibnamefont {{Viel}}},\ }\href {\doibase 10.1088/1475-7516/2014/12/053}
  {\bibfield  {journal} {\bibinfo  {journal} {\jcap}\ }\textbf {\bibinfo
  {volume} {2014}},\ \bibinfo {eid} {053} (\bibinfo {year} {2014})},\ \Eprint
  {http://arxiv.org/abs/1410.6813} {arXiv:1410.6813 [astro-ph.CO]} \BibitemShut
  {NoStop}%
\bibitem [{\citenamefont {{Liu}}\ \emph {et~al.}(2018)\citenamefont {{Liu}},
  \citenamefont {{Bird}},\ and\ \citenamefont {{Zorrilla Matilla et
  al.}}}]{2018JCAP...03..049L}%
  \BibitemOpen
  \bibfield  {author} {\bibinfo {author} {\bibfnamefont {J.}~\bibnamefont
  {{Liu}}}, \bibinfo {author} {\bibfnamefont {S.}~\bibnamefont {{Bird}}}, \
  and\ \bibinfo {author} {\bibfnamefont {J.~M.}\ \bibnamefont {{Zorrilla
  Matilla et al.}}},\ }\href {\doibase 10.1088/1475-7516/2018/03/049}
  {\bibfield  {journal} {\bibinfo  {journal} {\jcap}\ }\textbf {\bibinfo
  {volume} {2018}},\ \bibinfo {eid} {049} (\bibinfo {year} {2018})},\ \Eprint
  {http://arxiv.org/abs/1711.10524} {arXiv:1711.10524 [astro-ph.CO]}
  \BibitemShut {NoStop}%
\bibitem [{\citenamefont {{Navarro}}\ \emph {et~al.}(1997)\citenamefont
  {{Navarro}}, \citenamefont {{Frenk}},\ and\ \citenamefont
  {{White}}}]{1997ApJ...490..493N}%
  \BibitemOpen
  \bibfield  {author} {\bibinfo {author} {\bibfnamefont {J.~F.}\ \bibnamefont
  {{Navarro}}}, \bibinfo {author} {\bibfnamefont {C.~S.}\ \bibnamefont
  {{Frenk}}}, \ and\ \bibinfo {author} {\bibfnamefont {S.~D.~M.}\ \bibnamefont
  {{White}}},\ }\href {\doibase 10.1086/304888} {\bibfield  {journal} {\bibinfo
   {journal} {\apj}\ }\textbf {\bibinfo {volume} {490}},\ \bibinfo {pages}
  {493} (\bibinfo {year} {1997})},\ \Eprint
  {http://arxiv.org/abs/astro-ph/9611107} {arXiv:astro-ph/9611107 [astro-ph]}
  \BibitemShut {NoStop}%
\bibitem [{\citenamefont {{Villaescusa-Navarro}}\ \emph
  {et~al.}(2013)\citenamefont {{Villaescusa-Navarro}}, \citenamefont
  {{Vogelsberger}}, \citenamefont {{Viel}},\ and\ \citenamefont
  {{Loeb}}}]{2013MNRAS.431.3670V}%
  \BibitemOpen
  \bibfield  {author} {\bibinfo {author} {\bibfnamefont {F.}~\bibnamefont
  {{Villaescusa-Navarro}}}, \bibinfo {author} {\bibfnamefont {M.}~\bibnamefont
  {{Vogelsberger}}}, \bibinfo {author} {\bibfnamefont {M.}~\bibnamefont
  {{Viel}}}, \ and\ \bibinfo {author} {\bibfnamefont {A.}~\bibnamefont
  {{Loeb}}},\ }\href {\doibase 10.1093/mnras/stt452} {\bibfield  {journal}
  {\bibinfo  {journal} {\mnras}\ }\textbf {\bibinfo {volume} {431}},\ \bibinfo
  {pages} {3670} (\bibinfo {year} {2013})},\ \Eprint
  {http://arxiv.org/abs/1106.2543} {arXiv:1106.2543 [astro-ph.CO]} \BibitemShut
  {NoStop}%
\bibitem [{\citenamefont {{Massara}}\ \emph {et~al.}(2020)\citenamefont
  {{Massara}}, \citenamefont {{Villaescusa-Navarro}}, \citenamefont {{Ho}},
  \citenamefont {{Dalal}},\ and\ \citenamefont
  {{Spergel}}}]{2020arXiv200111024M}%
  \BibitemOpen
  \bibfield  {author} {\bibinfo {author} {\bibfnamefont {E.}~\bibnamefont
  {{Massara}}}, \bibinfo {author} {\bibfnamefont {F.}~\bibnamefont
  {{Villaescusa-Navarro}}}, \bibinfo {author} {\bibfnamefont {S.}~\bibnamefont
  {{Ho}}}, \bibinfo {author} {\bibfnamefont {N.}~\bibnamefont {{Dalal}}}, \
  and\ \bibinfo {author} {\bibfnamefont {D.~N.}\ \bibnamefont {{Spergel}}},\
  }\href@noop {} {\bibfield  {journal} {\bibinfo  {journal} {arXiv e-prints}\
  ,\ \bibinfo {eid} {arXiv:2001.11024}} (\bibinfo {year} {2020})},\ \Eprint
  {http://arxiv.org/abs/2001.11024} {arXiv:2001.11024 [astro-ph.CO]}
  \BibitemShut {NoStop}%
\bibitem [{\citenamefont {{Hamaus}}\ \emph {et~al.}(2010)\citenamefont
  {{Hamaus}}, \citenamefont {{Seljak}},\ and\ \citenamefont {{Desjacques et
  al.}}}]{2010PhRvD..82d3515H}%
  \BibitemOpen
  \bibfield  {author} {\bibinfo {author} {\bibfnamefont {N.}~\bibnamefont
  {{Hamaus}}}, \bibinfo {author} {\bibfnamefont {U.}~\bibnamefont {{Seljak}}},
  \ and\ \bibinfo {author} {\bibfnamefont {V.}~\bibnamefont {{Desjacques et
  al.}}},\ }\href {\doibase 10.1103/PhysRevD.82.043515} {\bibfield  {journal}
  {\bibinfo  {journal} {\prd}\ }\textbf {\bibinfo {volume} {82}},\ \bibinfo
  {eid} {043515} (\bibinfo {year} {2010})},\ \Eprint
  {http://arxiv.org/abs/1004.5377} {arXiv:1004.5377 [astro-ph.CO]} \BibitemShut
  {NoStop}%
\bibitem [{\citenamefont {{Zheng}}\ \emph {et~al.}(2005)\citenamefont
  {{Zheng}}, \citenamefont {{Berlind}},\ and\ \citenamefont {{Weinberg et
  al.}}}]{2005ApJ...633..791Z}%
  \BibitemOpen
  \bibfield  {author} {\bibinfo {author} {\bibfnamefont {Z.}~\bibnamefont
  {{Zheng}}}, \bibinfo {author} {\bibfnamefont {A.~A.}\ \bibnamefont
  {{Berlind}}}, \ and\ \bibinfo {author} {\bibfnamefont {D.~H.}\ \bibnamefont
  {{Weinberg et al.}}},\ }\href {\doibase 10.1086/466510} {\bibfield  {journal}
  {\bibinfo  {journal} {\apj}\ }\textbf {\bibinfo {volume} {633}},\ \bibinfo
  {pages} {791} (\bibinfo {year} {2005})},\ \Eprint
  {http://arxiv.org/abs/astro-ph/0408564} {arXiv:astro-ph/0408564 [astro-ph]}
  \BibitemShut {NoStop}%
\end{thebibliography}%


%

\end{document}